\documentclass[prl,twocolumn,twoside,preprintnumbers,superscriptaddress]{revtex4}
\usepackage{amsmath,slashed}
\usepackage{graphicx,graphics}
\usepackage{dcolumn}
\usepackage[hyperfootnotes=false]{hyperref}
\usepackage{xspace}

\newcommand{\GeV}{\,\text{GeV}}

\newcommand{\beq}{\begin{equation}}
\newcommand{\eeq}{\end{equation}}

\begin{document}

\preprint{INT-PUB-19-050}
\title{Short-distance constraints on hadronic light-by-light scattering \\[1mm] in the anomalous magnetic moment of the muon}

\author{Gilberto Colangelo}
\email[E-mail G.~Colangelo: ]{gilberto@itp.unibe.ch}
\affiliation{Albert Einstein Center for Fundamental Physics, Institute for Theoretical Physics, University of Bern, Sidlerstrasse 5, CH--3012 Bern, Switzerland}
\author{Franziska Hagelstein}
\email[E-mail F.~Hagelstein: ]{hagelstein@itp.unibe.ch}
\affiliation{Albert Einstein Center for Fundamental Physics, Institute for Theoretical Physics, University of Bern, Sidlerstrasse 5, CH--3012 Bern, Switzerland}
\author{Martin Hoferichter}
\email[E-mail M.~Hoferichter: ]{mhofer@uw.edu}
\affiliation{Institute for Nuclear Theory, University of Washington, Seattle, WA 98195-1550, USA}
\affiliation{Albert Einstein Center for Fundamental Physics, Institute for Theoretical Physics, University of Bern, Sidlerstrasse 5, CH--3012 Bern, Switzerland}
\author{Laetitia Laub}
\email[E-mail L.~Laub: ]{laub@itp.unibe.ch}
\affiliation{Albert Einstein Center for Fundamental Physics, Institute for Theoretical Physics, University of Bern, Sidlerstrasse 5, CH--3012 Bern, Switzerland}
\author{Peter Stoffer}
\email[E-mail P.~Stoffer: ]{pstoffer@ucsd.edu}
\affiliation{Department of Physics, University of California at San Diego, La Jolla, CA 92093, USA}

\begin{abstract}
  A key ingredient in the evaluation of hadronic light-by-light (HLbL)
  scattering in the anomalous magnetic moment of the muon $(g-2)_\mu$
  concerns short-distance constraints that follow from QCD by means
  of the operator product expansion.  Here we concentrate on the most
  important such constraint, in the longitudinal amplitudes, and show that
  it can be implemented efficiently in terms of a Regge sum over excited
  pseudoscalar states, constrained by phenomenological input on masses,
  two-photon couplings, as well as short-distance constraints on HLbL scattering and the
  pseudoscalar transition form factors.  Our estimate of the effect of the longitudinal short-distance constraints
  on the HLbL contribution is: $\Delta a_\mu^\text{LSDC}=13(6)\times
  10^{-11}$. This is significantly smaller than previous estimates, which
  mostly relied on an ad-hoc modification of the pseudoscalar poles
  and led to up to a $40\%$ increase with respect to the nominal
  pseudoscalar-pole contributions, when evaluated with modern input for the
  relevant transition form factors.  We also comment on the status of the
  transversal short-distance constraints and, by matching to perturbative QCD, argue that the
  corresponding correction will be significantly smaller than its
  longitudinal counterpart.
\end{abstract}

\maketitle

\section{Introduction}

The precision of the Standard-Model (SM) prediction for the anomalous magnetic moment of the muon, $a_\mu=(g-2)_\mu/2$, is limited by hadronic contributions. Already at the level of the 
current experiment~\cite{Bennett:2006fi}
\beq
a_\mu^\text{exp}=116\ 592\ 089(63)\times 10^{-11},
\eeq
estimates of the hadronic effects are crucial in evaluating the significance of the tension with the SM value, at the level of $3.5\,\sigma$.
With the forthcoming Fermilab E989 experiment~\cite{Grange:2015fou}, promising an improvement by a factor of $4$, as well as the 
E34 experiment at J-PARC~\cite{Abe:2019thb}, the SM model evaluation needs to follow suit. 

To this end, the relevant matrix elements need to be calculated either directly from QCD or be constrained by experimental data. 
The latter approach has traditionally been followed for hadronic vacuum polarization (HVP), which requires the two-point function of two electromagnetic currents 
and can be reconstructed from the cross section of $e^+e^-\to\text{hadrons}$~\cite{Keshavarzi:2018mgv,Jegerlehner:2018gjd,Colangelo:2018mtw,Hoferichter:2019gzf,Davier:2019can}. 
More recently, evaluations in lattice QCD have made significant progress~\cite{Borsanyi:2017zdw,Blum:2018mom,Giusti:2018mdh,Shintani:2019wai,Davies:2019efs,Gerardin:2019rua,Aubin:2019usy},
but are not yet at the level of the data-driven, dispersive approach. 

Next to HVP, the second-largest contribution to the uncertainty arises from hadronic light-by-light scattering. While also in this case progress in lattice QCD is promising~\cite{Blum:2016lnc,Blum:2017cer,Asmussen:2018oip}, another key development in recent years concerns the 
phenomenological evaluation, i.e., the use 
of dispersion relations to remove the reliance on hadronic models, 
either directly for the required four-point function that defines the HLbL tensor~\cite{Hoferichter:2013ama,Colangelo:2014dfa,Colangelo:2014pva,Colangelo:2015ama,Colangelo:2017qdm,Colangelo:2017fiz},
the Pauli form factor~\cite{Pauk:2014rfa}, or in terms of sum rules~\cite{Pascalutsa:2012pr,Green:2015sra,Danilkin:2016hnh,Hagelstein:2017obr,Hagelstein:2019tvp}.
In particular, organizing the calculation in terms of dispersion relations for the HLbL tensor has led to a solid 
understanding of the contributions related to the lowest-lying singularities---the single-particle poles from $P=\pi^0,\eta,\eta'$
and cuts from two-pion intermediate states---largely because the hadronic quantities determining the strength of these singularities, the 
$P\to\gamma^*\gamma^*$ transition form factors~\cite{Hoferichter:2014vra,Masjuan:2017tvw,Hoferichter:2018dmo,Hoferichter:2018kwz,Gerardin:2019vio,Eichmann:2019tjk}
and the helicity amplitudes
for $\gamma^*\gamma^*\to\pi\pi$~\cite{GarciaMartin:2010cw,Hoferichter:2011wk,Moussallam:2013una,Danilkin:2018qfn,Hoferichter:2019nlq,Danilkin:2019opj}, respectively,
can be provided as external input quantities, in a similar spirit as the $e^+e^-\to\text{hadrons}$ cross section for HVP. Higher-order iterations of HVP~\cite{Calmet:1976kd,Keshavarzi:2018mgv,Kurz:2014wya} and HLbL~\cite{Colangelo:2014qya} are already sufficiently under control.

For both HVP and HLbL, data-driven evaluations of the hadronic corrections to $(g-2)_\mu$ are fundamentally limited by the fact that experimental input is only available
in a given energy range, so that the tails of the dispersion integrals have to be estimated by other means, most notably short-distance constraints as derived from perturbative QCD (pQCD).  
In addition, even for HVP, short-distance constraints have been used for energies as low as
$2\GeV$ as a supplement to (and check of) experiment, 
with good agreement found between the pQCD prediction and data in between resonances~\cite{Keshavarzi:2018mgv,Davier:2019can}. For HLbL scattering such constraints become even more important
given the limited information on the HLbL tensor for intermediate and high energies. 

Two kinematic configurations are relevant for the HLbL contribution, one in which all photon virtualities $Q_i^2$ are large, and a second in which one of the non-vanishing virtualities remains 
small compared to the others $Q_3^2\ll Q_1^2\sim Q_2^2$.   
Recently, it was shown that the former situation can be addressed in a systematic operator product expansion (OPE), 
in which the pQCD quark loop emerges as the first term in the expansion~\cite{Bijnens:2019ghy}.
The second configuration is related to so-called mixed regions in the $g-2$ integral, i.e., integration regions in which asymptotic arguments only apply to 
a subset of the kinematic variables, while hadronic physics may still be relevant for others. 
A key insight derived in~\cite{Melnikov:2003xd} was that such effects can also be constrained with an OPE, by reducing the HLbL tensor to a vector--vector--axial-vector ($VVA$) 
three-point function and using known results for the corresponding anomaly and its (non-) renormalization~\cite{Vainshtein:2002nv,Knecht:2003xy,Knecht:2002hr,Czarnecki:2002nt,Jegerlehner:2005fs,Mondejar:2012sz}. 
The explicit implementation suggested in~\cite{Melnikov:2003xd} relied on the observation that both the OPE constraint and the normalization are satisfied
if the momentum dependence of the singly-virtual form factor describing the pseudoscalar-pole contribution is neglected. 
However, such a modification is not compatible with a description based on dispersion relations for the HLbL tensor. 

Here, we suggest to implement the corresponding longitudinal short-distance constraints in terms of excited pseudoscalar states. As we will show, not only can the asymptotic limits 
be implemented in a fairly economical manner, but the critical mixed regions can be constrained by phenomenological input for the masses and two-photon couplings
of the lowest pseudoscalar excitations. The model dependence can be further reduced by matching to the pQCD quark loop, which, in addition,
allows one to gain some insights into the scale where hadronic and pQCD-based descriptions should meet.

\section{OPE constraints on HLbL scattering}

The HLbL tensor is defined as the four-point function
\begin{align}
	\Pi^{\mu\nu\lambda\sigma}(q_1,q_2,q_3)&= -i \int d^4x \, d^4y \, d^4z \, e^{-i(q_1 \cdot x + q_2 \cdot y + q_3 \cdot z)} \notag\\
	&\hspace{-20pt}\times\langle 0 | T \{ j_\text{em}^\mu(x) j_\text{em}^\nu(y) j_\text{em}^\lambda(z) j_\text{em}^\sigma(0) \} | 0 \rangle
\end{align}
of four electromagnetic currents
\begin{align}
	j_\text{em}^\mu = \bar q \mathcal{Q} \gamma^\mu q, \quad \mathcal{Q} = \text{diag}\left(\frac{2}{3}, -\frac{1}{3}, -\frac{1}{3}\right),
\end{align}
where $q_i$ denote the photon virtualities, $q_4=q_1+q_2+q_3$, and $q = (u , d, s)^T$ the quark fields.
We work with the decomposition into scalar functions $\Pi_i$,
\beq
\Pi^{\mu\nu\lambda\sigma} = \sum_{i=1}^{54} T_i^{\mu\nu\lambda\sigma} \Pi_i,
\eeq
derived in~\cite{Colangelo:2015ama,Colangelo:2017fiz} following the general principle 
established by Bardeen, Tung~\cite{Bardeen:1969aw}, and Tarrach~\cite{Tarrach:1975tu}
(BTT). The contribution to $(g-2)_\mu$ then follows via
\begin{align}
a_\mu^\mathrm{HLbL} &= \frac{2 \alpha^3}{3 \pi^2} \int_0^\infty dQ_1 \int_0^\infty dQ_2 \int_{-1}^1 d\tau \sqrt{1-\tau^2}\, Q_1^3 Q_2^3 \notag\\
&\times\sum_{i=1}^{12} T_i(Q_1,Q_2,\tau) \bar \Pi_i(Q_1,Q_2,Q_3),
\end{align}
where $Q_i^2=-q_i^2$ are the Wick-rotated virtualities, $Q_3^2=Q_1^2+Q_2^2+2Q_1 Q_2\tau$, 
the $\bar \Pi_i$ refer to certain linear combinations of $\Pi_i$, and the $T_i$ are known kernel functions \cite{Colangelo:2015ama,Colangelo:2017fiz}.

In the limit where all $Q_i^2$ are large, 
the calculation from~\cite{Bijnens:2019ghy} proves the earlier statement of~\cite{Melnikov:2003xd} that the pQCD quark loop arises as the first term in a systematic OPE.
In particular, this implies the constraint
\beq
\label{pQCD}
 \lim_{Q\to\infty}Q^4\,\bar\Pi_1(Q,Q,Q)=-\frac{4}{9\pi^2}.
\eeq
The second kinematic configuration~\cite{Melnikov:2003xd}, $Q^2 \equiv Q_1^2\sim Q_2^2\gg Q_3^2$, when expressed in BTT basis, leads to the constraint
\beq
\label{MV}
\lim_{Q_3\to\infty}\lim_{Q\to\infty}Q^2Q_3^2\,\bar\Pi_1(Q,Q,Q_3)=-\frac{2}{3\pi^2}.
\eeq
The latter result can be derived by considering the $VVA$ triangle anomaly and its non-renormalization theorems~\cite{Vainshtein:2002nv,Knecht:2003xy,Knecht:2002hr,Czarnecki:2002nt,Jegerlehner:2005fs,Mondejar:2012sz}.
Its constraint on $\bar\Pi_1$ (and, by crossing symmetry, $\bar\Pi_2$) corresponds to the longitudinal amplitudes in the $VVA$ matrix element and we will therefore refer to 
$\bar\Pi_{1,2}$ as the longitudinal amplitudes and, accordingly, their constraints as longitudinal short-distance constraints.
Further, the limit~\eqref{MV} 
is intimately related to the pseudoscalar poles
\beq
\label{PS-pole}
\bar\Pi_1^{P\text{-pole}}=\frac{F_{P\gamma^*\gamma^*}(q_1^2,q_2^2)F_{P\gamma\gamma^*}(q_3^2)}{q_3^2-M_P^2},
\eeq
where $P=\pi^0,\eta,\eta'$, and the doubly-virtual $F_{P\gamma^*\gamma^*}(q_1^2,q_2^2)$ and singly-virtual $F_{P\gamma\gamma^*}(q_3^2)$ transition form factors determine the residue of the poles. They are subject to short-distance constraints themselves, for the pion we have the 
asymptotic constraint~\cite{Novikov:1983jt}
\beq
 \lim_{Q^2\rightarrow \infty} Q^2  F_{\pi^0 \gamma^*\gamma^*}(-Q^2,-Q^2)=\frac{2F_\pi}{3} 
 \label{Constraint1},
\eeq
as well as the Brodsky--Lepage limit~\cite{Lepage:1979zb,Lepage:1980fj,Brodsky:1981rp}
\beq
 \lim_{Q^2\rightarrow \infty} Q^2  F_{\pi^0 \gamma\gamma^*}(-Q^2)=2F_\pi. 
\label{Constraint2}
\eeq
Together with the normalization
\beq
F_{\pi^0\gamma\gamma}=\frac{1}{4\pi^2F_\pi},
\label{Constraint3}
\eeq
the former shows that if $F_{\pi^0\gamma\gamma^*}(q_3^2)\to F_{\pi^0\gamma\gamma}$ in~\eqref{PS-pole}, the pion decay constant $F_\pi$ would drop out 
and the pion would account for $-1/(6\pi^2)$ in~\eqref{MV}. Similarly, $\eta$ and $\eta'$ would provide the remaining $-1/(2\pi^2)$. This is 
the essence of the model suggested in~\cite{Melnikov:2003xd,Melnikov:2019xkq}. 

However, a constant singly-virtual transition form factor cannot be justified within a
dispersive approach for general HLbL scattering. Instead, one would need to
consider dispersion relations in the photon virtualities $q_i^2$ already in
reduced $g-2$ kinematics, and even then the residue would involve $F_{\pi^0
  \gamma\gamma^*}(M_P^2)$, not the normalization itself. Further, when
writing dispersion relations in the $q_i^2$ for $g-2$ kinematics, there is
no clear separation between the singularities of the HLbL amplitude and
those generated by hadronic intermediate states directly coupling to individual electromagnetic currents, such as $2 \pi$ states. In the dispersive
approach for general HLbL scattering the latter appear only in the transition form factors,
which factor out and can be treated as external input quantities. In this
sense, neglecting the momentum dependence of the singly-virtual transition form factor without
at the same time accounting for the additional cuts, leads to a distortion
of the low-energy properties of the HLbL tensor.

Instead, we propose here a solution based on a remark already made in~\cite{Melnikov:2003xd}: while a finite number of pseudoscalar poles, due to~\eqref{Constraint3},
cannot fulfill the OPE constraint~\eqref{MV}, an infinite series potentially can. The basic idea can be illustrated for large-$N_c$ Regge models of the transition form factor itself~\cite{RuizArriola:2006jge,Arriola:2010aq},
which assume a radial Regge trajectory to describe the masses of excited vector mesons,
\beq
M_{V(n)}^2=M_V^2+n\,\sigma_V^2,
\label{Regge}
\eeq
and rely on the ansatz:
\begin{align}
&F_{P \gamma^* \gamma^*}(-Q^2,-Q^2)\propto\sum_{n=0}^\infty \,\frac{1}{[Q^2+M_{V(n)}^2]^2}\notag\\
&=\frac{1}{\sigma_V^4} \, \psi ^{(1)} \left(\frac{M_V^2+Q^2}{\sigma_V^2}\right)\sim \frac{1}{Q^2},
\label{simpleExample}
\end{align}
with $\psi^{(1)}$ the trigamma function and the Regge slope $\sigma_V$.
In this way, the infinite sum produces the correct asymptotic behavior~\eqref{Constraint2}, even though none of the individual terms do.

One may wonder about the fate of the infinite sum over excited pseudoscalar states in the chiral limit, 
given that their decay constants are expected to vanish with the quark masses. We show below how the matching to pQCD 
removes the model dependence regarding which states are used to satisfy the short-distance constraints, 
so that the implementation in terms of pseudoscalar excitations mainly adds an estimate for the 
mixed-region contribution, driven by the phenomenology of the lowest excitations as well as 
the respective short-distance constraints.

\section{Large-$\boldsymbol{N_c}$ Regge model}

In the following, we present a large-$N_c$-inspired Regge model in the pseudoscalar and vector-meson sectors of QCD that allows us to satisfy the short-distance constraints
via an infinite sum of pseudoscalar-pole diagrams (see,
e.g.,~\cite{Peris:1998nj,Knecht:1998sp,Bijnens:2003rc} for the use of
large-$N_c$ arguments to simultaneously fulfill low- and high-energy
constraints). For brevity, we focus our description on the pion, referring
for a complete and more detailed account to~\cite{Colangelo:2019uex}.
We start from a standard large-$N_c$ ansatz for the pion transition form factors as in~\eqref{simpleExample}, 
but differentiate between $\rho$ and $\omega$ trajectories, which are assumed to enter with diagonal couplings due to  
the wave function overlap~\cite{RuizArriola:2006jge,Arriola:2010aq}. In a first step, we seek an extension of this model that 
satisfies the constraints~\eqref{Constraint1}--\eqref{Constraint3} for the transition form factor as well as~\eqref{pQCD} and~\eqref{MV} for the HLbL tensor
\begin{align}
&F_{\pi(n) \gamma^* \gamma^*}(-Q_1^2,-Q_2^2)\nonumber\\
&=\frac{1}{8 \pi^2
    F_\pi} \nonumber  \left\{ \left(\frac{M_\rho^2 M_\omega^2}{D_{\rho(n)}^1 D_{\omega(n)}^2} +
  \frac{M_\rho^2 M_\omega^2}{D_{\rho(n)}^2 D_{\omega(n)}^1}
\right)\Bigg[c_\mathrm{anom}
\right. \nonumber \\ 
 &+\left.\frac{1}{\Lambda^2}\left(
        c_A M_{+,\,n}^2+c_B M_{-,\,n}^2\right)+ c_\mathrm{diag} 
\frac{Q_1^2 Q_2^2}{\Lambda^2(Q_+^2+M^2_\mathrm{diag})} \right]\nonumber\\
&+\frac{Q_-^2}{Q_+^2} \left[c_\mathrm{BL}+\frac{1}{\Lambda^2}\left(
        c_A M_{-,\,n}^2+c_B M_{+,\,n}^2\right)  \right]\nonumber \\
 & \times\left.
\left(\frac{M_\rho^2 M_\omega^2}{D_{\rho(n)}^1 D_{\omega(n)}^2} -
  \frac{M_\rho^2 M_\omega^2}{D_{\rho(n)}^2 D_{\omega(n)}^1}
\right)\right\}, \label{TFFpi}
\end{align}
where $ Q_\pm^2=Q_1^2\pm Q_2^2$, $
M_{\pm,\,n}^2=\frac{1}{2} [M_{\omega(n)}^2\pm M_{\rho(n)}^2 ]$, $D^i_X=Q_i^2+M_X^2\label{def1}$,  
and $\Lambda = \mathcal{O}(1\, \text{GeV})$ a typical QCD scale. The five dimensionless parameters $c_\mathrm{anom}$, $c_A$, $c_B$, $c_\mathrm{diag}$, $c_\mathrm{BL}$
are used to fulfill all the constraints, while the remaining parameter $M_\mathrm{diag}$ is adjusted to reproduce the ground-state $\pi^0$ transition form factor~\cite{Hoferichter:2018dmo,Hoferichter:2018kwz}.
In the minimal model~\eqref{TFFpi}, we only allow $\pi(n)$ to couple to $\rho(n)$ and $\omega(n)$, i.e., the couplings are fully diagonal in the excitation numbers, while the 
effect of the eliminated vector-meson excitations is subsumed into a $Q_i^2$ dependence of the numerator multiplying the resonance propagators. 
In addition, we also considered an untruncated large-$N_c$ model, in which both the Regge summation in the transition form factor itself~\eqref{simpleExample}
and the HLbL tensor are retained, to assess the systematics in the large-$N_c$ ansatz~\cite{Colangelo:2019uex}.
Using the Regge slopes from~\cite{Masjuan:2012gc} and the other input parameters from~\cite{Tanabashi:2018oca}, we find that we can indeed reproduce well the $\pi^0$ transition form factor, 
which also ensures that effective-field-theory constraints on the pion-pole contribution to $(g-2)_\mu$~\cite{Knecht:2001qg,RamseyMusolf:2002cy} are fulfilled~\cite{Hoferichter:2018kwz}.
Finally, the model 
predicts a two-photon coupling of the first excited pion, $\pi(1300)$, in line with its
phenomenological bound $F_{\pi(1300)\gamma\gamma}<0.0544(71)\,\text{GeV}^{-1}$ \cite{Acciarri:1997rb,Salvini:2004gz}.

Constructing a large-$N_c$ Regge model for $\eta^{(\prime)}$ proceeds along the same lines, but involves several complications. 
First, the $\rho$ and $\omega$ trajectories do not suffice to incorporate all constraints since due to the $I=0$ nature of $\eta^{(\prime)}$ only equal-mass combinations of vector mesons ($2\rho$, $2\omega$, $2\phi$) contribute
to~\eqref{TFFpi}, so that only three model parameters survive. To provide sufficient freedom in satisfying all constraints the consideration of $\omega$--$\phi$ mixing cannot be avoided. 
In addition, $\eta$--$\eta'$ mixing needs to be taken into account, both for the flavor decomposition of the short-distance constraints as well as the relative weights of the vector-meson combinations in the transition form factors. 
The former is directly constrained by data on the transition form factors~\cite{Escribano:2015yup,Masjuan:2017tvw}, but for the calculation of the weights, which we extract 
from effective pseudoscalar--vector--vector and photon--vector Lagrangians~\cite{Landsberg:1986fd,Meissner:1987ge}, it is more convenient 
to work with the phenomenological two-angle mixing scheme
from~\cite{Feldmann:1998vh,Feldmann:1999uf}. We therefore use the latter
everywhere. All variants are covered by the error analysis.  

The resulting $\eta$ and $\eta'$ transition form factors are in good agreement with experimental data in the singly-virtual~\cite{Acciarri:1997yx,Behrend:1990sr,Gronberg:1997fj,BABAR:2011ad} and 
doubly-virtual regions~\cite{BaBar:2018zpn}, as well as the fit results using Canterbury approximants~\cite{Masjuan:2017tvw}. 
Furthermore, there are some phenomenological constraints on the two-photon couplings
for $\eta(1295)$ \cite{Acciarri:2000ev,Ahohe:2005ug}, $\eta(1405)$ \cite{Ahohe:2005ug,Acciarri:2000ev,Ablikim:2018ajr}, $\eta(1475)$ \cite{Achard:2007hm,Ablikim:2018ajr}, $\eta(1760)$ \cite{Zhang:2012tj}, and $X(1835)$ \cite{Zhang:2012tj,Ablikim:2018ajr}, where $\eta(1475)$ and $\eta(1760)$ are actually seen in $\gamma\gamma$ collisions, while for the others only limits are available. 
The detailed comparison depends on the assignment of these states into $\eta$ and $\eta'$ trajectories, but the predictions of our model 
are compatible with either the assignment from~\cite{Masjuan:2012gc,Tanabashi:2018oca} (our main choice) or the one from~\cite{Klempt:2007cp}, see~\cite{Colangelo:2019uex}.

By construction, the ground-state pseudoscalar-pole contributions to $(g-2)_\mu$ reproduce literature values~\cite{Hoferichter:2018dmo,Hoferichter:2018kwz,Masjuan:2017tvw,PabloSanchezPrivateCom} within errors, 
while the sum over excited-pseudoscalar poles leads to the increase:
\begin{align}
 \Delta a_\mu^{\pi\text{-poles}}&=2.7\,(0.4)_\text{Model}\,(1.2)_\text{syst}\times 10^{-11},\notag\\
  \Delta a_\mu^{\eta\text{-poles}}&=3.4^{\,+0.9}_{\,-0.7}\big\vert_\text{Model}\,(0.9)_\text{syst}\times 10^{-11},\notag \\
   \Delta a_\mu^{\eta'\text{-poles}}&=6.5\,(1.1)_\text{Model}\,(1.7)_\text{syst}\times 10^{-11},
\label{DeltaACombinedPionEtaP}
\end{align}
where the first error refers to the uncertainties propagated from the input parameters and the systematic error is estimated by comparison to an 
alternative untruncated large-$N_c$ Regge model~\cite{Colangelo:2019uex}. Combining all pseudoscalars, we find:
\begin{align}
 \Delta a_\mu^{\text{PS-poles}}&=\Delta a_\mu^{\pi\text{-poles}}+\Delta a_\mu^{\eta\text{-poles}}+\Delta a_\mu^{\eta'\text{-poles}}\notag\\
 &=12.6^{\,+1.6}_{\,-1.5}\big\vert_\text{Model}\,(3.8)_\text{syst}\times 10^{-11}\nonumber\\
 &=12.6(4.1)\times 10^{-11}.
 \label{final}
\end{align}
This result should be contrasted with the one suggested in~\cite{Melnikov:2003xd} to satisfy the mixed-region short-distance constraint (using transition form factor models from~\cite{Knecht:2001qf}): $\Delta a_\mu^{\text{PS-poles}}\big|_\text{MV}=23.5\times 10^{-11}$, which 
would become $38\times 10^{-11}$ once updated with modern input for the transition form factors, and thus suggest an increase nearly three times as large as~\eqref{final} or  
almost $40\%$ of the nominal pseudoscalar-pole contribution. 
Given that arguments following~\cite{Melnikov:2003xd} have been included in previous compilations of HLbL~\cite{Prades:2009tw}, a central result of this work is that 
such a large increase does not occur if the short-distance constraints are implemented without compromising the low-energy properties of HLbL scattering.

\section{Matching to perturbative QCD}

\begin{figure}[t]
 \centering
 \includegraphics[width=0.8\linewidth]{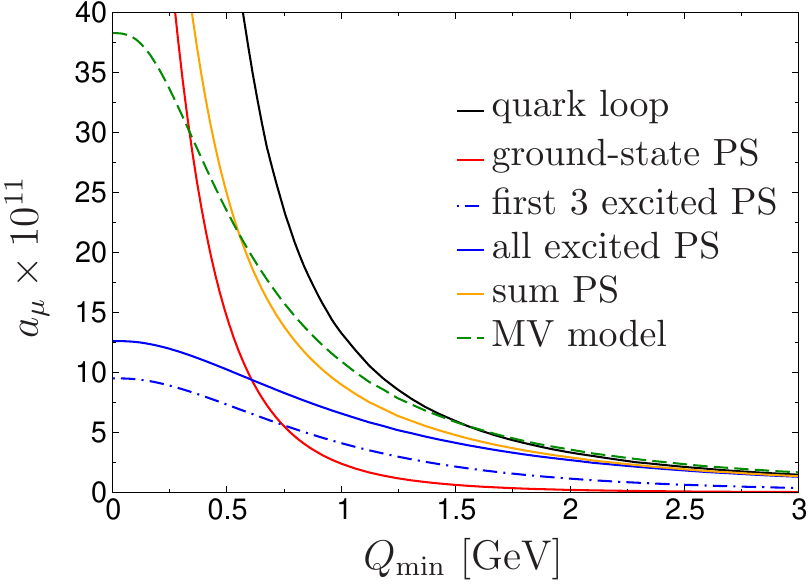}\\
 \vspace{0.4cm}
 \includegraphics[width=0.8\linewidth]{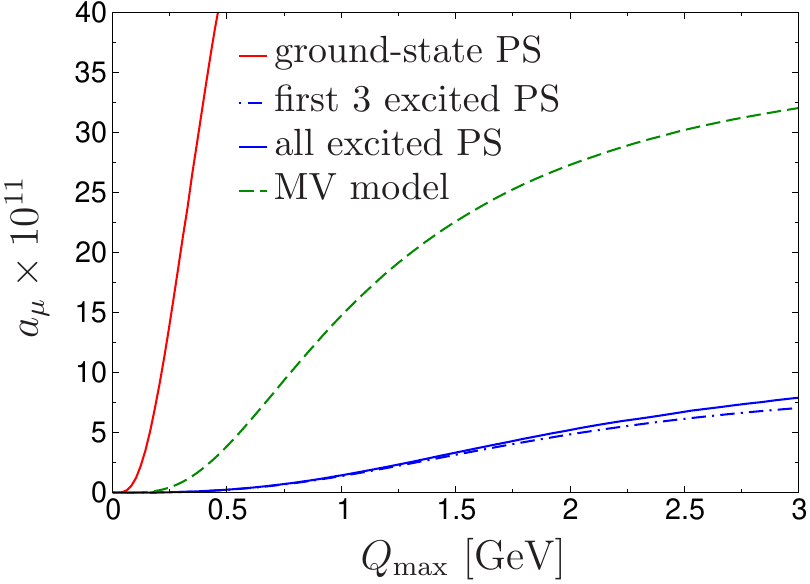}
 \caption{The longitudinal part of the massless pQCD quark loop (black), the ground-state pseudoscalars (red), the sum of all excitations from the large-$N_c$ Regge model (blue), the 
 first three excitations (blue dot-dashed), the sum of ground and excited states (orange),
	and the increase found in the Melnikov--Vainshtein model (green dashed). The upper plot shows the contribution to $a_\mu$ for $Q_{i}\geq Q_\text{min}$, the lower for $Q_{i}\leq Q_\text{max}$.}
 \label{fig:matching}
\end{figure}

Since, by construction, the sum over the pseudoscalar excitations fulfills the short-distance constraints, it has to match onto the pQCD quark loop for sufficiently large momentum transfers. In the upper plot of Fig.~\ref{fig:matching}, the contribution to $(g-2)_\mu$ from the massless pQCD quark loop (black) and the pseudoscalar-pole contributions (sum of ground-state and excited states in orange) are compared after imposing a cutoff $Q_\text{min}\leq Q_i$ in the integration: the matching occurs 
somewhere around $1.5$--$2\GeV$. The lower plot, where the opposite cutoff $Q_\text{max}\geq Q_i$ is imposed, shows that the contribution of the excited pseudoscalars (blue) to the low-energy region is very small and entirely saturated by the first few excitations (blue dot-dashed). 
These observations suggest to evaluate the asymptotic part of the integral $Q_i\geq Q_\text{min}$ with pQCD, to make explicit that this part of the result does not depend on the nature of hadronic states used in the implementation. Defining an optimal matching scale would require 
information on the uncertainty of the pQCD result. Here, we simply use a
rough $20\%$ estimate, which is the size of pQCD corrections for inclusive
$\tau$ decays, a process that has a similar energy scale and has been studied in detail~\cite{Davier:2008sk,Beneke:2008ad,Maltman:2008nf,Narison:2009vy,Caprini:2009vf,Pich:2013lsa}.

Together with the uncertainties from the Regge model, these considerations lead to a scale $Q_\text{match}=1.7\GeV$. Varying this scale within $\pm 0.5\GeV$ 
and adding the systematic uncertainty from the comparison to the untruncated Regge model, we obtain as our final result:
\begin{align}
 \Delta a_\mu^\text{LSDC}&=\left[8.7(5.5)_{\text{PS-poles}}+4.6(9)_{\text{pQCD}}\right]\times 10^{-11} \notag\\
 &= 13(6)\times 10^{-11}
\label{final_result}
\end{align}
for the increase of $(g-2)_\mu$ due to longitudinal short-distance constraints. 
In particular, the lowest three pseudoscalar excitations, whose contribution is at least 
partly constrained by phenomenological input on masses and two-photon couplings, give $7.8\times 10^{-11}$. Given that the most uncertain contribution, from $n>3$, thus amounts to only $10\%$
of the total, the uncertainty estimate in~\eqref{final_result} should be conservative enough to cover the remaining model dependence. 
In particular, the error in~\eqref{final_result} includes an inflation of the Regge slope uncertainties
by a factor three, to allow for systematic effects that might occur if other hadronic states were used to implement the short-distance constraints.  
More recently, this expectation has been confirmed by models in holographic QCD based on a summation of an infinite tower of axial-vector resonances instead~\cite{Leutgeb:2019gbz,Cappiello:2019hwh}, which despite very different assumptions and systematics yield results remarkably close to~\eqref{final_result}.

\begin{figure}[t]
 \centering
 \includegraphics[width=0.8\linewidth]{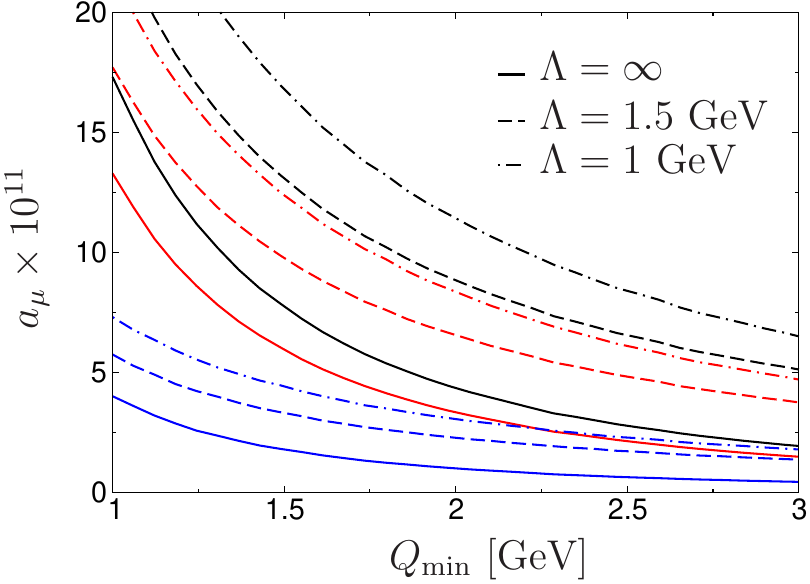}
 \caption{Contribution of the massless pQCD quark loop to $a_\mu$ from the region $Q_{1,2}\geq Q_\text{min}$, with the contribution from $Q_3$ below $Q_\text{min}$ damped by $Q_3^2/(Q_3^2+\Lambda^2)$ (plus crossed). 
 The total contribution (black) is split into longitudinal (red) and transversal (blue) components. The limit $Q_i\geq Q_\text{min}$ for all $Q_i$ is reproduced for $\Lambda\to\infty$.}
 \label{fig:quark_loop}
\end{figure}

With the impact of the longitudinal short-distance constraints estimated as in~\eqref{final_result}, it is natural to inquire about the role of the transversal short-distance constraints. A first estimate could again be obtained 
by matching to pQCD. Fig.~\ref{fig:quark_loop} extends the integration region beyond $Q_i\geq Q_\text{min}$ into the mixed region, but suppressing this additional contribution by a
factor $Q_3^2/(Q_3^2+\Lambda^2)$, because otherwise part of the ground-state pseudoscalar contribution would be double-counted. 
The longitudinal result is reproduced for scales around $\Lambda\sim Q_\text{min}\sim 1.4\GeV$, for which one would read off $\Delta a_\mu^\text{TSDC}\sim 4\times 10^{-11}$. 
Accordingly, we would expect the impact of the transversal short-distance constraints to be significantly less than that of the longitudinal ones. 

We stress that the calculation presented here is complementary 
to higher-order calculations in pQCD and/or the OPE~\cite{Bijnens:2019ghy}, which would allow one to improve the matching between hadronic implementations and a perturbative description.
Similarly, more experimental guidance on the two-photon couplings of hadronic states in the $1$--$2\GeV$ region would be beneficial for 
the phenomenological analysis, not only for the excited pseudoscalars, but for axial-vector resonances as well, which outlines avenues for future work. 
We conclude that with the present analysis the biggest systematic uncertainty due to short-distance constraints has been reduced significantly, 
with the result that the asymptotic part of the HLbL tensor is under sufficient 
control for the first release from the Fermilab experiment.

\section*{Acknowledgments}
\begin{acknowledgments}
We thank R.~Arnaldi, P.~Bickert, J.~Bijnens, S.~Eidelman, A.~G\'{e}rardin, C.~Hanhart, S.~Holz, B.~Kubis, S.~Leupold, J.~L{\"u}dtke, A.~Manohar, V.~Metag, M.~Procura, 
E.~Ruiz Arriola, P.~Sanchez-Puertas, A.~Uras, G.~Usai, A.~Vainshtein, and E.~Weil for useful communication on various aspects of this work. 
Financial support by  the DOE (Grant Nos.\ DE-FG02-00ER41132 and DE-SC0009919) and the Swiss National Science Foundation
is gratefully acknowledged. M.H.\ is supported by an SNSF Eccellenza Professorial Fellowship (Project No.\ PCEFP2\_181117).
\end{acknowledgments}

\bibliography{AMM}

\begin{thebibliography}{99}
\expandafter\ifx\csname natexlab\endcsname\relax\def\natexlab#1{#1}\fi
\expandafter\ifx\csname bibnamefont\endcsname\relax
  \def\bibnamefont#1{#1}\fi
\expandafter\ifx\csname bibfnamefont\endcsname\relax
  \def\bibfnamefont#1{#1}\fi
\expandafter\ifx\csname citenamefont\endcsname\relax
  \def\citenamefont#1{#1}\fi
\expandafter\ifx\csname url\endcsname\relax
  \def\url#1{\texttt{#1}}\fi
\expandafter\ifx\csname urlprefix\endcsname\relax\def\urlprefix{URL }\fi
\providecommand{\bibinfo}[2]{#2}
\providecommand{\eprint}[2][]{\url{#2}}

\bibitem[{\citenamefont{Bennett et~al.}(2006)}]{Bennett:2006fi}
\bibinfo{author}{\bibfnamefont{G.~W.} \bibnamefont{Bennett}}
  \bibnamefont{et~al.} (\bibinfo{collaboration}{Muon $g-2$}),
  \bibinfo{journal}{Phys. Rev.} \textbf{\bibinfo{volume}{D 73}},
  \bibinfo{pages}{072003} (\bibinfo{year}{2006}), \eprint{hep-ex/0602035}.

\bibitem[{\citenamefont{Grange et~al.}(2015)}]{Grange:2015fou}
\bibinfo{author}{\bibfnamefont{J.}~\bibnamefont{Grange}} \bibnamefont{et~al.}
  (\bibinfo{collaboration}{Muon $g-2$}) (\bibinfo{year}{2015}),
  \eprint{1501.06858}.

\bibitem[{\citenamefont{Abe et~al.}(2019)}]{Abe:2019thb}
\bibinfo{author}{\bibfnamefont{M.}~\bibnamefont{Abe}} \bibnamefont{et~al.},
  \bibinfo{journal}{PTEP} \textbf{\bibinfo{volume}{2019}},
  \bibinfo{pages}{053C02} (\bibinfo{year}{2019}), \eprint{1901.03047}.

\bibitem[{\citenamefont{Keshavarzi et~al.}(2018)\citenamefont{Keshavarzi,
  Nomura, and Teubner}}]{Keshavarzi:2018mgv}
\bibinfo{author}{\bibfnamefont{A.}~\bibnamefont{Keshavarzi}},
  \bibinfo{author}{\bibfnamefont{D.}~\bibnamefont{Nomura}}, \bibnamefont{and}
  \bibinfo{author}{\bibfnamefont{T.}~\bibnamefont{Teubner}},
  \bibinfo{journal}{Phys. Rev.} \textbf{\bibinfo{volume}{D 97}},
  \bibinfo{pages}{114025} (\bibinfo{year}{2018}), \eprint{1802.02995}.

\bibitem[{\citenamefont{Jegerlehner}(2019)}]{Jegerlehner:2018gjd}
\bibinfo{author}{\bibfnamefont{F.}~\bibnamefont{Jegerlehner}},
  \bibinfo{journal}{EPJ Web Conf.} \textbf{\bibinfo{volume}{199}},
  \bibinfo{pages}{01010} (\bibinfo{year}{2019}), \eprint{1809.07413}.

\bibitem[{\citenamefont{Colangelo
  et~al.}(2019{\natexlab{a}})\citenamefont{Colangelo, Hoferichter, and
  Stoffer}}]{Colangelo:2018mtw}
\bibinfo{author}{\bibfnamefont{G.}~\bibnamefont{Colangelo}},
  \bibinfo{author}{\bibfnamefont{M.}~\bibnamefont{Hoferichter}},
  \bibnamefont{and} \bibinfo{author}{\bibfnamefont{P.}~\bibnamefont{Stoffer}},
  \bibinfo{journal}{JHEP} \textbf{\bibinfo{volume}{02}}, \bibinfo{pages}{006}
  (\bibinfo{year}{2019}{\natexlab{a}}), \eprint{1810.00007}.

\bibitem[{\citenamefont{Hoferichter et~al.}(2019)\citenamefont{Hoferichter,
  Hoid, and Kubis}}]{Hoferichter:2019gzf}
\bibinfo{author}{\bibfnamefont{M.}~\bibnamefont{Hoferichter}},
  \bibinfo{author}{\bibfnamefont{B.-L.} \bibnamefont{Hoid}}, \bibnamefont{and}
  \bibinfo{author}{\bibfnamefont{B.}~\bibnamefont{Kubis}},
  \bibinfo{journal}{JHEP} \textbf{\bibinfo{volume}{08}}, \bibinfo{pages}{137}
  (\bibinfo{year}{2019}), \eprint{1907.01556}.

\bibitem[{\citenamefont{Davier et~al.}(2019)\citenamefont{Davier, Hoecker,
  Malaescu, and Zhang}}]{Davier:2019can}
\bibinfo{author}{\bibfnamefont{M.}~\bibnamefont{Davier}},
  \bibinfo{author}{\bibfnamefont{A.}~\bibnamefont{Hoecker}},
  \bibinfo{author}{\bibfnamefont{B.}~\bibnamefont{Malaescu}}, \bibnamefont{and}
  \bibinfo{author}{\bibfnamefont{Z.}~\bibnamefont{Zhang}}
  (\bibinfo{year}{2019}), \eprint{1908.00921}.

\bibitem[{\citenamefont{Borsanyi et~al.}(2018)}]{Borsanyi:2017zdw}
\bibinfo{author}{\bibfnamefont{S.}~\bibnamefont{Borsanyi}} \bibnamefont{et~al.}
  (\bibinfo{collaboration}{Budapest-Marseille-Wuppertal}),
  \bibinfo{journal}{Phys. Rev. Lett.} \textbf{\bibinfo{volume}{121}},
  \bibinfo{pages}{022002} (\bibinfo{year}{2018}), \eprint{1711.04980}.

\bibitem[{\citenamefont{Blum et~al.}(2018)\citenamefont{Blum, Boyle,
  G{\"u}lpers, Izubuchi, Jin, Jung, J{\"u}ttner, Lehner, Portelli, and
  Tsang}}]{Blum:2018mom}
\bibinfo{author}{\bibfnamefont{T.}~\bibnamefont{Blum}},
  \bibinfo{author}{\bibfnamefont{P.~A.} \bibnamefont{Boyle}},
  \bibinfo{author}{\bibfnamefont{V.}~\bibnamefont{G{\"u}lpers}},
  \bibinfo{author}{\bibfnamefont{T.}~\bibnamefont{Izubuchi}},
  \bibinfo{author}{\bibfnamefont{L.}~\bibnamefont{Jin}},
  \bibinfo{author}{\bibfnamefont{C.}~\bibnamefont{Jung}},
  \bibinfo{author}{\bibfnamefont{A.}~\bibnamefont{J{\"u}ttner}},
  \bibinfo{author}{\bibfnamefont{C.}~\bibnamefont{Lehner}},
  \bibinfo{author}{\bibfnamefont{A.}~\bibnamefont{Portelli}}, \bibnamefont{and}
  \bibinfo{author}{\bibfnamefont{J.~T.} \bibnamefont{Tsang}}
  (\bibinfo{collaboration}{RBC, UKQCD}), \bibinfo{journal}{Phys. Rev. Lett.}
  \textbf{\bibinfo{volume}{121}}, \bibinfo{pages}{022003}
  (\bibinfo{year}{2018}), \eprint{1801.07224}.

\bibitem[{\citenamefont{Giusti et~al.}(2018)\citenamefont{Giusti, Sanfilippo,
  and Simula}}]{Giusti:2018mdh}
\bibinfo{author}{\bibfnamefont{D.}~\bibnamefont{Giusti}},
  \bibinfo{author}{\bibfnamefont{F.}~\bibnamefont{Sanfilippo}},
  \bibnamefont{and} \bibinfo{author}{\bibfnamefont{S.}~\bibnamefont{Simula}},
  \bibinfo{journal}{Phys. Rev.} \textbf{\bibinfo{volume}{D 98}},
  \bibinfo{pages}{114504} (\bibinfo{year}{2018}), \eprint{1808.00887}.

\bibitem[{\citenamefont{Shintani and Kuramashi}(2019)}]{Shintani:2019wai}
\bibinfo{author}{\bibfnamefont{E.}~\bibnamefont{Shintani}} \bibnamefont{and}
  \bibinfo{author}{\bibfnamefont{Y.}~\bibnamefont{Kuramashi}}
  (\bibinfo{collaboration}{PACS}), \bibinfo{journal}{Phys. Rev.}
  \textbf{\bibinfo{volume}{D 100}}, \bibinfo{pages}{034517}
  (\bibinfo{year}{2019}), \eprint{1902.00885}.

\bibitem[{\citenamefont{Davies et~al.}(2020)}]{Davies:2019efs}
\bibinfo{author}{\bibfnamefont{C.~T.~H.} \bibnamefont{Davies}}
  \bibnamefont{et~al.} (\bibinfo{collaboration}{Fermilab Lattice,
  LATTICE-HPQCD, MILC}), \bibinfo{journal}{Phys. Rev.}
  \textbf{\bibinfo{volume}{D101}}, \bibinfo{pages}{034512}
  (\bibinfo{year}{2020}), \eprint{1902.04223}.

\bibitem[{\citenamefont{G{\'e}rardin
  et~al.}(2019{\natexlab{a}})\citenamefont{G{\'e}rardin, C{\`e}, von Hippel,
  H{\"o}rz, Meyer, Mohler, Ottnad, Wilhelm, and Wittig}}]{Gerardin:2019rua}
\bibinfo{author}{\bibfnamefont{A.}~\bibnamefont{G{\'e}rardin}},
  \bibinfo{author}{\bibfnamefont{M.}~\bibnamefont{C{\`e}}},
  \bibinfo{author}{\bibfnamefont{G.}~\bibnamefont{von Hippel}},
  \bibinfo{author}{\bibfnamefont{B.}~\bibnamefont{H{\"o}rz}},
  \bibinfo{author}{\bibfnamefont{H.~B.} \bibnamefont{Meyer}},
  \bibinfo{author}{\bibfnamefont{D.}~\bibnamefont{Mohler}},
  \bibinfo{author}{\bibfnamefont{K.}~\bibnamefont{Ottnad}},
  \bibinfo{author}{\bibfnamefont{J.}~\bibnamefont{Wilhelm}}, \bibnamefont{and}
  \bibinfo{author}{\bibfnamefont{H.}~\bibnamefont{Wittig}},
  \bibinfo{journal}{Phys. Rev.} \textbf{\bibinfo{volume}{D 100}},
  \bibinfo{pages}{014510} (\bibinfo{year}{2019}{\natexlab{a}}),
  \eprint{1904.03120}.

\bibitem[{\citenamefont{Aubin et~al.}(2020)\citenamefont{Aubin, Blum, Tu,
  Golterman, Jung, and Peris}}]{Aubin:2019usy}
\bibinfo{author}{\bibfnamefont{C.}~\bibnamefont{Aubin}},
  \bibinfo{author}{\bibfnamefont{T.}~\bibnamefont{Blum}},
  \bibinfo{author}{\bibfnamefont{C.}~\bibnamefont{Tu}},
  \bibinfo{author}{\bibfnamefont{M.}~\bibnamefont{Golterman}},
  \bibinfo{author}{\bibfnamefont{C.}~\bibnamefont{Jung}}, \bibnamefont{and}
  \bibinfo{author}{\bibfnamefont{S.}~\bibnamefont{Peris}},
  \bibinfo{journal}{Phys. Rev.} \textbf{\bibinfo{volume}{D101}},
  \bibinfo{pages}{014503} (\bibinfo{year}{2020}), \eprint{1905.09307}.

\bibitem[{\citenamefont{Blum et~al.}(2017{\natexlab{a}})\citenamefont{Blum,
  Christ, Hayakawa, Izubuchi, Jin, Jung, and Lehner}}]{Blum:2016lnc}
\bibinfo{author}{\bibfnamefont{T.}~\bibnamefont{Blum}},
  \bibinfo{author}{\bibfnamefont{N.}~\bibnamefont{Christ}},
  \bibinfo{author}{\bibfnamefont{M.}~\bibnamefont{Hayakawa}},
  \bibinfo{author}{\bibfnamefont{T.}~\bibnamefont{Izubuchi}},
  \bibinfo{author}{\bibfnamefont{L.}~\bibnamefont{Jin}},
  \bibinfo{author}{\bibfnamefont{C.}~\bibnamefont{Jung}}, \bibnamefont{and}
  \bibinfo{author}{\bibfnamefont{C.}~\bibnamefont{Lehner}},
  \bibinfo{journal}{Phys. Rev. Lett.} \textbf{\bibinfo{volume}{118}},
  \bibinfo{pages}{022005} (\bibinfo{year}{2017}{\natexlab{a}}),
  \eprint{1610.04603}.

\bibitem[{\citenamefont{Blum et~al.}(2017{\natexlab{b}})\citenamefont{Blum,
  Christ, Hayakawa, Izubuchi, Jin, Jung, and Lehner}}]{Blum:2017cer}
\bibinfo{author}{\bibfnamefont{T.}~\bibnamefont{Blum}},
  \bibinfo{author}{\bibfnamefont{N.}~\bibnamefont{Christ}},
  \bibinfo{author}{\bibfnamefont{M.}~\bibnamefont{Hayakawa}},
  \bibinfo{author}{\bibfnamefont{T.}~\bibnamefont{Izubuchi}},
  \bibinfo{author}{\bibfnamefont{L.}~\bibnamefont{Jin}},
  \bibinfo{author}{\bibfnamefont{C.}~\bibnamefont{Jung}}, \bibnamefont{and}
  \bibinfo{author}{\bibfnamefont{C.}~\bibnamefont{Lehner}},
  \bibinfo{journal}{Phys. Rev.} \textbf{\bibinfo{volume}{D 96}},
  \bibinfo{pages}{034515} (\bibinfo{year}{2017}{\natexlab{b}}),
  \eprint{1705.01067}.

\bibitem[{\citenamefont{Asmussen et~al.}(2019)\citenamefont{Asmussen,
  G{\'e}rardin, Nyffeler, and Meyer}}]{Asmussen:2018oip}
\bibinfo{author}{\bibfnamefont{N.}~\bibnamefont{Asmussen}},
  \bibinfo{author}{\bibfnamefont{A.}~\bibnamefont{G{\'e}rardin}},
  \bibinfo{author}{\bibfnamefont{A.}~\bibnamefont{Nyffeler}}, \bibnamefont{and}
  \bibinfo{author}{\bibfnamefont{H.~B.} \bibnamefont{Meyer}},
  \bibinfo{journal}{SciPost Phys. Proc.} \textbf{\bibinfo{volume}{1}},
  \bibinfo{pages}{031} (\bibinfo{year}{2019}), \eprint{1811.08320}.

\bibitem[{\citenamefont{Hoferichter
  et~al.}(2014{\natexlab{a}})\citenamefont{Hoferichter, Colangelo, Procura, and
  Stoffer}}]{Hoferichter:2013ama}
\bibinfo{author}{\bibfnamefont{M.}~\bibnamefont{Hoferichter}},
  \bibinfo{author}{\bibfnamefont{G.}~\bibnamefont{Colangelo}},
  \bibinfo{author}{\bibfnamefont{M.}~\bibnamefont{Procura}}, \bibnamefont{and}
  \bibinfo{author}{\bibfnamefont{P.}~\bibnamefont{Stoffer}},
  \bibinfo{journal}{Int. J. Mod. Phys. Conf. Ser.}
  \textbf{\bibinfo{volume}{35}}, \bibinfo{pages}{1460400}
  (\bibinfo{year}{2014}{\natexlab{a}}), \eprint{1309.6877}.

\bibitem[{\citenamefont{Colangelo
  et~al.}(2014{\natexlab{a}})\citenamefont{Colangelo, Hoferichter, Procura, and
  Stoffer}}]{Colangelo:2014dfa}
\bibinfo{author}{\bibfnamefont{G.}~\bibnamefont{Colangelo}},
  \bibinfo{author}{\bibfnamefont{M.}~\bibnamefont{Hoferichter}},
  \bibinfo{author}{\bibfnamefont{M.}~\bibnamefont{Procura}}, \bibnamefont{and}
  \bibinfo{author}{\bibfnamefont{P.}~\bibnamefont{Stoffer}},
  \bibinfo{journal}{JHEP} \textbf{\bibinfo{volume}{09}}, \bibinfo{pages}{091}
  (\bibinfo{year}{2014}{\natexlab{a}}), \eprint{1402.7081}.

\bibitem[{\citenamefont{Colangelo
  et~al.}(2014{\natexlab{b}})\citenamefont{Colangelo, Hoferichter, Kubis,
  Procura, and Stoffer}}]{Colangelo:2014pva}
\bibinfo{author}{\bibfnamefont{G.}~\bibnamefont{Colangelo}},
  \bibinfo{author}{\bibfnamefont{M.}~\bibnamefont{Hoferichter}},
  \bibinfo{author}{\bibfnamefont{B.}~\bibnamefont{Kubis}},
  \bibinfo{author}{\bibfnamefont{M.}~\bibnamefont{Procura}}, \bibnamefont{and}
  \bibinfo{author}{\bibfnamefont{P.}~\bibnamefont{Stoffer}},
  \bibinfo{journal}{Phys. Lett.} \textbf{\bibinfo{volume}{B 738}},
  \bibinfo{pages}{6} (\bibinfo{year}{2014}{\natexlab{b}}), \eprint{1408.2517}.

\bibitem[{\citenamefont{Colangelo et~al.}(2015)\citenamefont{Colangelo,
  Hoferichter, Procura, and Stoffer}}]{Colangelo:2015ama}
\bibinfo{author}{\bibfnamefont{G.}~\bibnamefont{Colangelo}},
  \bibinfo{author}{\bibfnamefont{M.}~\bibnamefont{Hoferichter}},
  \bibinfo{author}{\bibfnamefont{M.}~\bibnamefont{Procura}}, \bibnamefont{and}
  \bibinfo{author}{\bibfnamefont{P.}~\bibnamefont{Stoffer}},
  \bibinfo{journal}{JHEP} \textbf{\bibinfo{volume}{09}}, \bibinfo{pages}{074}
  (\bibinfo{year}{2015}), \eprint{1506.01386}.

\bibitem[{\citenamefont{Colangelo
  et~al.}(2017{\natexlab{a}})\citenamefont{Colangelo, Hoferichter, Procura, and
  Stoffer}}]{Colangelo:2017qdm}
\bibinfo{author}{\bibfnamefont{G.}~\bibnamefont{Colangelo}},
  \bibinfo{author}{\bibfnamefont{M.}~\bibnamefont{Hoferichter}},
  \bibinfo{author}{\bibfnamefont{M.}~\bibnamefont{Procura}}, \bibnamefont{and}
  \bibinfo{author}{\bibfnamefont{P.}~\bibnamefont{Stoffer}},
  \bibinfo{journal}{Phys. Rev. Lett.} \textbf{\bibinfo{volume}{118}},
  \bibinfo{pages}{232001} (\bibinfo{year}{2017}{\natexlab{a}}),
  \eprint{1701.06554}.

\bibitem[{\citenamefont{Colangelo
  et~al.}(2017{\natexlab{b}})\citenamefont{Colangelo, Hoferichter, Procura, and
  Stoffer}}]{Colangelo:2017fiz}
\bibinfo{author}{\bibfnamefont{G.}~\bibnamefont{Colangelo}},
  \bibinfo{author}{\bibfnamefont{M.}~\bibnamefont{Hoferichter}},
  \bibinfo{author}{\bibfnamefont{M.}~\bibnamefont{Procura}}, \bibnamefont{and}
  \bibinfo{author}{\bibfnamefont{P.}~\bibnamefont{Stoffer}},
  \bibinfo{journal}{JHEP} \textbf{\bibinfo{volume}{04}}, \bibinfo{pages}{161}
  (\bibinfo{year}{2017}{\natexlab{b}}), \eprint{1702.07347}.

\bibitem[{\citenamefont{Pauk and Vanderhaeghen}(2014)}]{Pauk:2014rfa}
\bibinfo{author}{\bibfnamefont{V.}~\bibnamefont{Pauk}} \bibnamefont{and}
  \bibinfo{author}{\bibfnamefont{M.}~\bibnamefont{Vanderhaeghen}},
  \bibinfo{journal}{Phys. Rev.} \textbf{\bibinfo{volume}{D 90}},
  \bibinfo{pages}{113012} (\bibinfo{year}{2014}), \eprint{1409.0819}.

\bibitem[{\citenamefont{Pascalutsa et~al.}(2012)\citenamefont{Pascalutsa, Pauk,
  and Vanderhaeghen}}]{Pascalutsa:2012pr}
\bibinfo{author}{\bibfnamefont{V.}~\bibnamefont{Pascalutsa}},
  \bibinfo{author}{\bibfnamefont{V.}~\bibnamefont{Pauk}}, \bibnamefont{and}
  \bibinfo{author}{\bibfnamefont{M.}~\bibnamefont{Vanderhaeghen}},
  \bibinfo{journal}{Phys. Rev.} \textbf{\bibinfo{volume}{D 85}},
  \bibinfo{pages}{116001} (\bibinfo{year}{2012}), \eprint{1204.0740}.

\bibitem[{\citenamefont{Green et~al.}(2015)\citenamefont{Green, Gryniuk, von
  Hippel, Meyer, and Pascalutsa}}]{Green:2015sra}
\bibinfo{author}{\bibfnamefont{J.}~\bibnamefont{Green}},
  \bibinfo{author}{\bibfnamefont{O.}~\bibnamefont{Gryniuk}},
  \bibinfo{author}{\bibfnamefont{G.}~\bibnamefont{von Hippel}},
  \bibinfo{author}{\bibfnamefont{H.~B.} \bibnamefont{Meyer}}, \bibnamefont{and}
  \bibinfo{author}{\bibfnamefont{V.}~\bibnamefont{Pascalutsa}},
  \bibinfo{journal}{Phys. Rev. Lett.} \textbf{\bibinfo{volume}{115}},
  \bibinfo{pages}{222003} (\bibinfo{year}{2015}), \eprint{1507.01577}.

\bibitem[{\citenamefont{Danilkin and Vanderhaeghen}(2017)}]{Danilkin:2016hnh}
\bibinfo{author}{\bibfnamefont{I.}~\bibnamefont{Danilkin}} \bibnamefont{and}
  \bibinfo{author}{\bibfnamefont{M.}~\bibnamefont{Vanderhaeghen}},
  \bibinfo{journal}{Phys. Rev.} \textbf{\bibinfo{volume}{D 95}},
  \bibinfo{pages}{014019} (\bibinfo{year}{2017}), \eprint{1611.04646}.

\bibitem[{\citenamefont{Hagelstein and Pascalutsa}(2018)}]{Hagelstein:2017obr}
\bibinfo{author}{\bibfnamefont{F.}~\bibnamefont{Hagelstein}} \bibnamefont{and}
  \bibinfo{author}{\bibfnamefont{V.}~\bibnamefont{Pascalutsa}},
  \bibinfo{journal}{Phys. Rev. Lett.} \textbf{\bibinfo{volume}{120}},
  \bibinfo{pages}{072002} (\bibinfo{year}{2018}), \eprint{1710.04571}.

\bibitem[{\citenamefont{Hagelstein and Pascalutsa}(2019)}]{Hagelstein:2019tvp}
\bibinfo{author}{\bibfnamefont{F.}~\bibnamefont{Hagelstein}} \bibnamefont{and}
  \bibinfo{author}{\bibfnamefont{V.}~\bibnamefont{Pascalutsa}}
  (\bibinfo{year}{2019}), \eprint{1907.06927}.

\bibitem[{\citenamefont{Hoferichter
  et~al.}(2014{\natexlab{b}})\citenamefont{Hoferichter, Kubis, Leupold,
  Niecknig, and Schneider}}]{Hoferichter:2014vra}
\bibinfo{author}{\bibfnamefont{M.}~\bibnamefont{Hoferichter}},
  \bibinfo{author}{\bibfnamefont{B.}~\bibnamefont{Kubis}},
  \bibinfo{author}{\bibfnamefont{S.}~\bibnamefont{Leupold}},
  \bibinfo{author}{\bibfnamefont{F.}~\bibnamefont{Niecknig}}, \bibnamefont{and}
  \bibinfo{author}{\bibfnamefont{S.~P.} \bibnamefont{Schneider}},
  \bibinfo{journal}{Eur. Phys. J.} \textbf{\bibinfo{volume}{C 74}},
  \bibinfo{pages}{3180} (\bibinfo{year}{2014}{\natexlab{b}}),
  \eprint{1410.4691}.

\bibitem[{\citenamefont{Masjuan and Sanchez-Puertas}(2017)}]{Masjuan:2017tvw}
\bibinfo{author}{\bibfnamefont{P.}~\bibnamefont{Masjuan}} \bibnamefont{and}
  \bibinfo{author}{\bibfnamefont{P.}~\bibnamefont{Sanchez-Puertas}},
  \bibinfo{journal}{Phys. Rev.} \textbf{\bibinfo{volume}{D 95}},
  \bibinfo{pages}{054026} (\bibinfo{year}{2017}), \eprint{1701.05829}.

\bibitem[{\citenamefont{Hoferichter
  et~al.}(2018{\natexlab{a}})\citenamefont{Hoferichter, Hoid, Kubis, Leupold,
  and Schneider}}]{Hoferichter:2018dmo}
\bibinfo{author}{\bibfnamefont{M.}~\bibnamefont{Hoferichter}},
  \bibinfo{author}{\bibfnamefont{B.-L.} \bibnamefont{Hoid}},
  \bibinfo{author}{\bibfnamefont{B.}~\bibnamefont{Kubis}},
  \bibinfo{author}{\bibfnamefont{S.}~\bibnamefont{Leupold}}, \bibnamefont{and}
  \bibinfo{author}{\bibfnamefont{S.~P.} \bibnamefont{Schneider}},
  \bibinfo{journal}{Phys. Rev. Lett.} \textbf{\bibinfo{volume}{121}},
  \bibinfo{pages}{112002} (\bibinfo{year}{2018}{\natexlab{a}}),
  \eprint{1805.01471}.

\bibitem[{\citenamefont{Hoferichter
  et~al.}(2018{\natexlab{b}})\citenamefont{Hoferichter, Hoid, Kubis, Leupold,
  and Schneider}}]{Hoferichter:2018kwz}
\bibinfo{author}{\bibfnamefont{M.}~\bibnamefont{Hoferichter}},
  \bibinfo{author}{\bibfnamefont{B.-L.} \bibnamefont{Hoid}},
  \bibinfo{author}{\bibfnamefont{B.}~\bibnamefont{Kubis}},
  \bibinfo{author}{\bibfnamefont{S.}~\bibnamefont{Leupold}}, \bibnamefont{and}
  \bibinfo{author}{\bibfnamefont{S.~P.} \bibnamefont{Schneider}},
  \bibinfo{journal}{JHEP} \textbf{\bibinfo{volume}{10}}, \bibinfo{pages}{141}
  (\bibinfo{year}{2018}{\natexlab{b}}), \eprint{1808.04823}.

\bibitem[{\citenamefont{G{\'e}rardin
  et~al.}(2019{\natexlab{b}})\citenamefont{G{\'e}rardin, Meyer, and
  Nyffeler}}]{Gerardin:2019vio}
\bibinfo{author}{\bibfnamefont{A.}~\bibnamefont{G{\'e}rardin}},
  \bibinfo{author}{\bibfnamefont{H.~B.} \bibnamefont{Meyer}}, \bibnamefont{and}
  \bibinfo{author}{\bibfnamefont{A.}~\bibnamefont{Nyffeler}},
  \bibinfo{journal}{Phys. Rev.} \textbf{\bibinfo{volume}{D 100}},
  \bibinfo{pages}{034520} (\bibinfo{year}{2019}{\natexlab{b}}),
  \eprint{1903.09471}.

\bibitem[{\citenamefont{Eichmann et~al.}(2019)\citenamefont{Eichmann, Fischer,
  Weil, and Williams}}]{Eichmann:2019tjk}
\bibinfo{author}{\bibfnamefont{G.}~\bibnamefont{Eichmann}},
  \bibinfo{author}{\bibfnamefont{C.~S.} \bibnamefont{Fischer}},
  \bibinfo{author}{\bibfnamefont{E.}~\bibnamefont{Weil}}, \bibnamefont{and}
  \bibinfo{author}{\bibfnamefont{R.}~\bibnamefont{Williams}},
  \bibinfo{journal}{Phys. Lett.} \textbf{\bibinfo{volume}{B 797}},
  \bibinfo{pages}{134855} (\bibinfo{year}{2019}), \eprint{1903.10844}.

\bibitem[{\citenamefont{Garc{\'i}a-Mart{\'i}n and
  Moussallam}(2010)}]{GarciaMartin:2010cw}
\bibinfo{author}{\bibfnamefont{R.}~\bibnamefont{Garc{\'i}a-Mart{\'i}n}}
  \bibnamefont{and}
  \bibinfo{author}{\bibfnamefont{B.}~\bibnamefont{Moussallam}},
  \bibinfo{journal}{Eur. Phys. J.} \textbf{\bibinfo{volume}{C 70}},
  \bibinfo{pages}{155} (\bibinfo{year}{2010}), \eprint{1006.5373}.

\bibitem[{\citenamefont{Hoferichter et~al.}(2011)\citenamefont{Hoferichter,
  Phillips, and Schat}}]{Hoferichter:2011wk}
\bibinfo{author}{\bibfnamefont{M.}~\bibnamefont{Hoferichter}},
  \bibinfo{author}{\bibfnamefont{D.~R.} \bibnamefont{Phillips}},
  \bibnamefont{and} \bibinfo{author}{\bibfnamefont{C.}~\bibnamefont{Schat}},
  \bibinfo{journal}{Eur. Phys. J.} \textbf{\bibinfo{volume}{C 71}},
  \bibinfo{pages}{1743} (\bibinfo{year}{2011}), \eprint{1106.4147}.

\bibitem[{\citenamefont{Moussallam}(2013)}]{Moussallam:2013una}
\bibinfo{author}{\bibfnamefont{B.}~\bibnamefont{Moussallam}},
  \bibinfo{journal}{Eur. Phys. J.} \textbf{\bibinfo{volume}{C 73}},
  \bibinfo{pages}{2539} (\bibinfo{year}{2013}), \eprint{1305.3143}.

\bibitem[{\citenamefont{Danilkin and Vanderhaeghen}(2019)}]{Danilkin:2018qfn}
\bibinfo{author}{\bibfnamefont{I.}~\bibnamefont{Danilkin}} \bibnamefont{and}
  \bibinfo{author}{\bibfnamefont{M.}~\bibnamefont{Vanderhaeghen}},
  \bibinfo{journal}{Phys. Lett.} \textbf{\bibinfo{volume}{B 789}},
  \bibinfo{pages}{366} (\bibinfo{year}{2019}), \eprint{1810.03669}.

\bibitem[{\citenamefont{Hoferichter and Stoffer}(2019)}]{Hoferichter:2019nlq}
\bibinfo{author}{\bibfnamefont{M.}~\bibnamefont{Hoferichter}} \bibnamefont{and}
  \bibinfo{author}{\bibfnamefont{P.}~\bibnamefont{Stoffer}},
  \bibinfo{journal}{JHEP} \textbf{\bibinfo{volume}{07}}, \bibinfo{pages}{073}
  (\bibinfo{year}{2019}), \eprint{1905.13198}.

\bibitem[{\citenamefont{Danilkin et~al.}(2020)\citenamefont{Danilkin, Deineka,
  and Vanderhaeghen}}]{Danilkin:2019opj}
\bibinfo{author}{\bibfnamefont{I.}~\bibnamefont{Danilkin}},
  \bibinfo{author}{\bibfnamefont{O.}~\bibnamefont{Deineka}}, \bibnamefont{and}
  \bibinfo{author}{\bibfnamefont{M.}~\bibnamefont{Vanderhaeghen}},
  \bibinfo{journal}{Phys. Rev.} \textbf{\bibinfo{volume}{D101}},
  \bibinfo{pages}{054008} (\bibinfo{year}{2020}), \eprint{1909.04158}.

\bibitem[{\citenamefont{Calmet et~al.}(1976)\citenamefont{Calmet, Narison,
  Perrottet, and de~Rafael}}]{Calmet:1976kd}
\bibinfo{author}{\bibfnamefont{J.}~\bibnamefont{Calmet}},
  \bibinfo{author}{\bibfnamefont{S.}~\bibnamefont{Narison}},
  \bibinfo{author}{\bibfnamefont{M.}~\bibnamefont{Perrottet}},
  \bibnamefont{and}
  \bibinfo{author}{\bibfnamefont{E.}~\bibnamefont{de~Rafael}},
  \bibinfo{journal}{Phys. Lett.} \textbf{\bibinfo{volume}{61B}},
  \bibinfo{pages}{283} (\bibinfo{year}{1976}).

\bibitem[{\citenamefont{Kurz et~al.}(2014)\citenamefont{Kurz, Liu, Marquard,
  and Steinhauser}}]{Kurz:2014wya}
\bibinfo{author}{\bibfnamefont{A.}~\bibnamefont{Kurz}},
  \bibinfo{author}{\bibfnamefont{T.}~\bibnamefont{Liu}},
  \bibinfo{author}{\bibfnamefont{P.}~\bibnamefont{Marquard}}, \bibnamefont{and}
  \bibinfo{author}{\bibfnamefont{M.}~\bibnamefont{Steinhauser}},
  \bibinfo{journal}{Phys. Lett.} \textbf{\bibinfo{volume}{B734}},
  \bibinfo{pages}{144} (\bibinfo{year}{2014}), \eprint{1403.6400}.

\bibitem[{\citenamefont{Colangelo
  et~al.}(2014{\natexlab{c}})\citenamefont{Colangelo, Hoferichter, Nyffeler,
  Passera, and Stoffer}}]{Colangelo:2014qya}
\bibinfo{author}{\bibfnamefont{G.}~\bibnamefont{Colangelo}},
  \bibinfo{author}{\bibfnamefont{M.}~\bibnamefont{Hoferichter}},
  \bibinfo{author}{\bibfnamefont{A.}~\bibnamefont{Nyffeler}},
  \bibinfo{author}{\bibfnamefont{M.}~\bibnamefont{Passera}}, \bibnamefont{and}
  \bibinfo{author}{\bibfnamefont{P.}~\bibnamefont{Stoffer}},
  \bibinfo{journal}{Phys. Lett.} \textbf{\bibinfo{volume}{B735}},
  \bibinfo{pages}{90} (\bibinfo{year}{2014}{\natexlab{c}}), \eprint{1403.7512}.

\bibitem[{\citenamefont{Bijnens et~al.}(2019)\citenamefont{Bijnens,
  Hermansson-Truedsson, and Rodr{\'i}guez-S{\'a}nchez}}]{Bijnens:2019ghy}
\bibinfo{author}{\bibfnamefont{J.}~\bibnamefont{Bijnens}},
  \bibinfo{author}{\bibfnamefont{N.}~\bibnamefont{Hermansson-Truedsson}},
  \bibnamefont{and}
  \bibinfo{author}{\bibfnamefont{A.}~\bibnamefont{Rodr{\'i}guez-S{\'a}nchez}},
  \bibinfo{journal}{Phys. Lett.} \textbf{\bibinfo{volume}{B798}},
  \bibinfo{pages}{134994} (\bibinfo{year}{2019}), \eprint{1908.03331}.

\bibitem[{\citenamefont{Melnikov and Vainshtein}(2004)}]{Melnikov:2003xd}
\bibinfo{author}{\bibfnamefont{K.}~\bibnamefont{Melnikov}} \bibnamefont{and}
  \bibinfo{author}{\bibfnamefont{A.}~\bibnamefont{Vainshtein}},
  \bibinfo{journal}{Phys. Rev.} \textbf{\bibinfo{volume}{D 70}},
  \bibinfo{pages}{113006} (\bibinfo{year}{2004}), \eprint{hep-ph/0312226}.

\bibitem[{\citenamefont{Vainshtein}(2003)}]{Vainshtein:2002nv}
\bibinfo{author}{\bibfnamefont{A.}~\bibnamefont{Vainshtein}},
  \bibinfo{journal}{Phys. Lett.} \textbf{\bibinfo{volume}{B 569}},
  \bibinfo{pages}{187} (\bibinfo{year}{2003}), \eprint{hep-ph/0212231}.

\bibitem[{\citenamefont{Knecht et~al.}(2004)\citenamefont{Knecht, Peris,
  Perrottet, and de~Rafael}}]{Knecht:2003xy}
\bibinfo{author}{\bibfnamefont{M.}~\bibnamefont{Knecht}},
  \bibinfo{author}{\bibfnamefont{S.}~\bibnamefont{Peris}},
  \bibinfo{author}{\bibfnamefont{M.}~\bibnamefont{Perrottet}},
  \bibnamefont{and}
  \bibinfo{author}{\bibfnamefont{E.}~\bibnamefont{de~Rafael}},
  \bibinfo{journal}{JHEP} \textbf{\bibinfo{volume}{03}}, \bibinfo{pages}{035}
  (\bibinfo{year}{2004}), \eprint{hep-ph/0311100}.

\bibitem[{\citenamefont{Knecht et~al.}(2002{\natexlab{a}})\citenamefont{Knecht,
  Peris, Perrottet, and de~Rafael}}]{Knecht:2002hr}
\bibinfo{author}{\bibfnamefont{M.}~\bibnamefont{Knecht}},
  \bibinfo{author}{\bibfnamefont{S.}~\bibnamefont{Peris}},
  \bibinfo{author}{\bibfnamefont{M.}~\bibnamefont{Perrottet}},
  \bibnamefont{and}
  \bibinfo{author}{\bibfnamefont{E.}~\bibnamefont{de~Rafael}},
  \bibinfo{journal}{JHEP} \textbf{\bibinfo{volume}{11}}, \bibinfo{pages}{003}
  (\bibinfo{year}{2002}{\natexlab{a}}), \eprint{hep-ph/0205102}.

\bibitem[{\citenamefont{Czarnecki et~al.}(2003)\citenamefont{Czarnecki,
  Marciano, and Vainshtein}}]{Czarnecki:2002nt}
\bibinfo{author}{\bibfnamefont{A.}~\bibnamefont{Czarnecki}},
  \bibinfo{author}{\bibfnamefont{W.~J.} \bibnamefont{Marciano}},
  \bibnamefont{and}
  \bibinfo{author}{\bibfnamefont{A.}~\bibnamefont{Vainshtein}},
  \bibinfo{journal}{Phys. Rev.} \textbf{\bibinfo{volume}{D 67}},
  \bibinfo{pages}{073006} (\bibinfo{year}{2003}), \bibinfo{note}{[Erratum:
  Phys. Rev. {\bf D 73}, 119901 (2006)]}, \eprint{hep-ph/0212229}.

\bibitem[{\citenamefont{Jegerlehner and Tarasov}(2006)}]{Jegerlehner:2005fs}
\bibinfo{author}{\bibfnamefont{F.}~\bibnamefont{Jegerlehner}} \bibnamefont{and}
  \bibinfo{author}{\bibfnamefont{O.~V.} \bibnamefont{Tarasov}},
  \bibinfo{journal}{Phys. Lett.} \textbf{\bibinfo{volume}{B 639}},
  \bibinfo{pages}{299} (\bibinfo{year}{2006}), \eprint{hep-ph/0510308}.

\bibitem[{\citenamefont{Mondejar and Melnikov}(2013)}]{Mondejar:2012sz}
\bibinfo{author}{\bibfnamefont{J.}~\bibnamefont{Mondejar}} \bibnamefont{and}
  \bibinfo{author}{\bibfnamefont{K.}~\bibnamefont{Melnikov}},
  \bibinfo{journal}{Phys. Lett.} \textbf{\bibinfo{volume}{B 718}},
  \bibinfo{pages}{1364} (\bibinfo{year}{2013}), \eprint{1210.0812}.

\bibitem[{\citenamefont{Bardeen and Tung}(1968)}]{Bardeen:1969aw}
\bibinfo{author}{\bibfnamefont{W.~A.} \bibnamefont{Bardeen}} \bibnamefont{and}
  \bibinfo{author}{\bibfnamefont{W.~K.} \bibnamefont{Tung}},
  \bibinfo{journal}{Phys. Rev.} \textbf{\bibinfo{volume}{173}},
  \bibinfo{pages}{1423} (\bibinfo{year}{1968}), \bibinfo{note}{[Erratum: Phys.
  Rev. {\bf D 4}, 3229 (1971)]}.

\bibitem[{\citenamefont{Tarrach}(1975)}]{Tarrach:1975tu}
\bibinfo{author}{\bibfnamefont{R.}~\bibnamefont{Tarrach}},
  \bibinfo{journal}{Nuovo Cim.} \textbf{\bibinfo{volume}{A 28}},
  \bibinfo{pages}{409} (\bibinfo{year}{1975}).

\bibitem[{\citenamefont{Novikov et~al.}(1984)\citenamefont{Novikov, Shifman,
  Vainshtein, Voloshin, and Zakharov}}]{Novikov:1983jt}
\bibinfo{author}{\bibfnamefont{V.~A.} \bibnamefont{Novikov}},
  \bibinfo{author}{\bibfnamefont{M.~A.} \bibnamefont{Shifman}},
  \bibinfo{author}{\bibfnamefont{A.~I.} \bibnamefont{Vainshtein}},
  \bibinfo{author}{\bibfnamefont{M.~B.} \bibnamefont{Voloshin}},
  \bibnamefont{and} \bibinfo{author}{\bibfnamefont{V.~I.}
  \bibnamefont{Zakharov}}, \bibinfo{journal}{Nucl. Phys.}
  \textbf{\bibinfo{volume}{B 237}}, \bibinfo{pages}{525}
  (\bibinfo{year}{1984}).

\bibitem[{\citenamefont{Lepage and Brodsky}(1979)}]{Lepage:1979zb}
\bibinfo{author}{\bibfnamefont{G.~P.} \bibnamefont{Lepage}} \bibnamefont{and}
  \bibinfo{author}{\bibfnamefont{S.~J.} \bibnamefont{Brodsky}},
  \bibinfo{journal}{Phys. Lett.} \textbf{\bibinfo{volume}{87B}},
  \bibinfo{pages}{359} (\bibinfo{year}{1979}).

\bibitem[{\citenamefont{Lepage and Brodsky}(1980)}]{Lepage:1980fj}
\bibinfo{author}{\bibfnamefont{G.~P.} \bibnamefont{Lepage}} \bibnamefont{and}
  \bibinfo{author}{\bibfnamefont{S.~J.} \bibnamefont{Brodsky}},
  \bibinfo{journal}{Phys. Rev.} \textbf{\bibinfo{volume}{D 22}},
  \bibinfo{pages}{2157} (\bibinfo{year}{1980}).

\bibitem[{\citenamefont{Brodsky and Lepage}(1981)}]{Brodsky:1981rp}
\bibinfo{author}{\bibfnamefont{S.~J.} \bibnamefont{Brodsky}} \bibnamefont{and}
  \bibinfo{author}{\bibfnamefont{G.~P.} \bibnamefont{Lepage}},
  \bibinfo{journal}{Phys. Rev.} \textbf{\bibinfo{volume}{D 24}},
  \bibinfo{pages}{1808} (\bibinfo{year}{1981}).

\bibitem[{\citenamefont{Melnikov and Vainshtein}(2019)}]{Melnikov:2019xkq}
\bibinfo{author}{\bibfnamefont{K.}~\bibnamefont{Melnikov}} \bibnamefont{and}
  \bibinfo{author}{\bibfnamefont{A.}~\bibnamefont{Vainshtein}}
  (\bibinfo{year}{2019}), \eprint{1911.05874}.

\bibitem[{\citenamefont{Ruiz~Arriola and
  Broniowski}(2006)}]{RuizArriola:2006jge}
\bibinfo{author}{\bibfnamefont{E.}~\bibnamefont{Ruiz~Arriola}}
  \bibnamefont{and}
  \bibinfo{author}{\bibfnamefont{W.}~\bibnamefont{Broniowski}},
  \bibinfo{journal}{Phys. Rev.} \textbf{\bibinfo{volume}{D 74}},
  \bibinfo{pages}{034008} (\bibinfo{year}{2006}), \eprint{hep-ph/0605318}.

\bibitem[{\citenamefont{Ruiz~Arriola and Broniowski}(2010)}]{Arriola:2010aq}
\bibinfo{author}{\bibfnamefont{E.}~\bibnamefont{Ruiz~Arriola}}
  \bibnamefont{and}
  \bibinfo{author}{\bibfnamefont{W.}~\bibnamefont{Broniowski}},
  \bibinfo{journal}{Phys. Rev.} \textbf{\bibinfo{volume}{D 81}},
  \bibinfo{pages}{094021} (\bibinfo{year}{2010}), \eprint{1004.0837}.

\bibitem[{\citenamefont{Peris et~al.}(1998)\citenamefont{Peris, Perrottet, and
  de~Rafael}}]{Peris:1998nj}
\bibinfo{author}{\bibfnamefont{S.}~\bibnamefont{Peris}},
  \bibinfo{author}{\bibfnamefont{M.}~\bibnamefont{Perrottet}},
  \bibnamefont{and}
  \bibinfo{author}{\bibfnamefont{E.}~\bibnamefont{de~Rafael}},
  \bibinfo{journal}{JHEP} \textbf{\bibinfo{volume}{05}}, \bibinfo{pages}{011}
  (\bibinfo{year}{1998}), \eprint{hep-ph/9805442}.

\bibitem[{\citenamefont{Knecht et~al.}(1998)\citenamefont{Knecht, Peris, and
  de~Rafael}}]{Knecht:1998sp}
\bibinfo{author}{\bibfnamefont{M.}~\bibnamefont{Knecht}},
  \bibinfo{author}{\bibfnamefont{S.}~\bibnamefont{Peris}}, \bibnamefont{and}
  \bibinfo{author}{\bibfnamefont{E.}~\bibnamefont{de~Rafael}},
  \bibinfo{journal}{Phys. Lett.} \textbf{\bibinfo{volume}{B 443}},
  \bibinfo{pages}{255} (\bibinfo{year}{1998}), \eprint{hep-ph/9809594}.

\bibitem[{\citenamefont{Bijnens et~al.}(2003)\citenamefont{Bijnens, G{\'a}miz,
  Lipartia, and Prades}}]{Bijnens:2003rc}
\bibinfo{author}{\bibfnamefont{J.}~\bibnamefont{Bijnens}},
  \bibinfo{author}{\bibfnamefont{E.}~\bibnamefont{G{\'a}miz}},
  \bibinfo{author}{\bibfnamefont{E.}~\bibnamefont{Lipartia}}, \bibnamefont{and}
  \bibinfo{author}{\bibfnamefont{J.}~\bibnamefont{Prades}},
  \bibinfo{journal}{JHEP} \textbf{\bibinfo{volume}{04}}, \bibinfo{pages}{055}
  (\bibinfo{year}{2003}), \eprint{hep-ph/0304222}.

\bibitem[{\citenamefont{Colangelo
  et~al.}(2019{\natexlab{b}})\citenamefont{Colangelo, Hagelstein, Hoferichter,
  Laub, and Stoffer}}]{Colangelo:2019uex}
\bibinfo{author}{\bibfnamefont{G.}~\bibnamefont{Colangelo}},
  \bibinfo{author}{\bibfnamefont{F.}~\bibnamefont{Hagelstein}},
  \bibinfo{author}{\bibfnamefont{M.}~\bibnamefont{Hoferichter}},
  \bibinfo{author}{\bibfnamefont{L.}~\bibnamefont{Laub}}, \bibnamefont{and}
  \bibinfo{author}{\bibfnamefont{P.}~\bibnamefont{Stoffer}}
  (\bibinfo{year}{2019}{\natexlab{b}}), \eprint{1910.13432}.

\bibitem[{\citenamefont{Masjuan et~al.}(2012)\citenamefont{Masjuan,
  Ruiz~Arriola, and Broniowski}}]{Masjuan:2012gc}
\bibinfo{author}{\bibfnamefont{P.}~\bibnamefont{Masjuan}},
  \bibinfo{author}{\bibfnamefont{E.}~\bibnamefont{Ruiz~Arriola}},
  \bibnamefont{and}
  \bibinfo{author}{\bibfnamefont{W.}~\bibnamefont{Broniowski}},
  \bibinfo{journal}{Phys. Rev.} \textbf{\bibinfo{volume}{D 85}},
  \bibinfo{pages}{094006} (\bibinfo{year}{2012}), \eprint{1203.4782}.

\bibitem[{\citenamefont{Tanabashi et~al.}(2018)}]{Tanabashi:2018oca}
\bibinfo{author}{\bibfnamefont{M.}~\bibnamefont{Tanabashi}}
  \bibnamefont{et~al.} (\bibinfo{collaboration}{Particle Data Group}),
  \bibinfo{journal}{Phys. Rev.} \textbf{\bibinfo{volume}{D 98}},
  \bibinfo{pages}{030001} (\bibinfo{year}{2018}).

\bibitem[{\citenamefont{Knecht et~al.}(2002{\natexlab{b}})\citenamefont{Knecht,
  Nyffeler, Perrottet, and de~Rafael}}]{Knecht:2001qg}
\bibinfo{author}{\bibfnamefont{M.}~\bibnamefont{Knecht}},
  \bibinfo{author}{\bibfnamefont{A.}~\bibnamefont{Nyffeler}},
  \bibinfo{author}{\bibfnamefont{M.}~\bibnamefont{Perrottet}},
  \bibnamefont{and}
  \bibinfo{author}{\bibfnamefont{E.}~\bibnamefont{de~Rafael}},
  \bibinfo{journal}{Phys. Rev. Lett.} \textbf{\bibinfo{volume}{88}},
  \bibinfo{pages}{071802} (\bibinfo{year}{2002}{\natexlab{b}}),
  \eprint{hep-ph/0111059}.

\bibitem[{\citenamefont{Ramsey-Musolf and Wise}(2002)}]{RamseyMusolf:2002cy}
\bibinfo{author}{\bibfnamefont{M.~J.} \bibnamefont{Ramsey-Musolf}}
  \bibnamefont{and} \bibinfo{author}{\bibfnamefont{M.~B.} \bibnamefont{Wise}},
  \bibinfo{journal}{Phys. Rev. Lett.} \textbf{\bibinfo{volume}{89}},
  \bibinfo{pages}{041601} (\bibinfo{year}{2002}), \eprint{hep-ph/0201297}.

\bibitem[{\citenamefont{Acciarri et~al.}(1997)}]{Acciarri:1997rb}
\bibinfo{author}{\bibfnamefont{M.}~\bibnamefont{Acciarri}} \bibnamefont{et~al.}
  (\bibinfo{collaboration}{L3}), \bibinfo{journal}{Phys. Lett.}
  \textbf{\bibinfo{volume}{B 413}}, \bibinfo{pages}{147}
  (\bibinfo{year}{1997}).

\bibitem[{\citenamefont{Salvini et~al.}(2004)}]{Salvini:2004gz}
\bibinfo{author}{\bibfnamefont{P.}~\bibnamefont{Salvini}} \bibnamefont{et~al.}
  (\bibinfo{collaboration}{OBELIX}), \bibinfo{journal}{Eur. Phys. J.}
  \textbf{\bibinfo{volume}{C 35}}, \bibinfo{pages}{21} (\bibinfo{year}{2004}).

\bibitem[{\citenamefont{Escribano et~al.}(2016)\citenamefont{Escribano,
  Gonz{\`a}lez-Sol{\'i}s, Masjuan, and Sanchez-Puertas}}]{Escribano:2015yup}
\bibinfo{author}{\bibfnamefont{R.}~\bibnamefont{Escribano}},
  \bibinfo{author}{\bibfnamefont{S.}~\bibnamefont{Gonz{\`a}lez-Sol{\'i}s}},
  \bibinfo{author}{\bibfnamefont{P.}~\bibnamefont{Masjuan}}, \bibnamefont{and}
  \bibinfo{author}{\bibfnamefont{P.}~\bibnamefont{Sanchez-Puertas}},
  \bibinfo{journal}{Phys. Rev.} \textbf{\bibinfo{volume}{D 94}},
  \bibinfo{pages}{054033} (\bibinfo{year}{2016}), \eprint{1512.07520}.

\bibitem[{\citenamefont{Landsberg}(1985)}]{Landsberg:1986fd}
\bibinfo{author}{\bibfnamefont{L.~G.} \bibnamefont{Landsberg}},
  \bibinfo{journal}{Phys. Rept.} \textbf{\bibinfo{volume}{128}},
  \bibinfo{pages}{301} (\bibinfo{year}{1985}).

\bibitem[{\citenamefont{Mei{\ss}ner}(1988)}]{Meissner:1987ge}
\bibinfo{author}{\bibfnamefont{U.-G.} \bibnamefont{Mei{\ss}ner}},
  \bibinfo{journal}{Phys. Rept.} \textbf{\bibinfo{volume}{161}},
  \bibinfo{pages}{213} (\bibinfo{year}{1988}).

\bibitem[{\citenamefont{Feldmann et~al.}(1998)\citenamefont{Feldmann, Kroll,
  and Stech}}]{Feldmann:1998vh}
\bibinfo{author}{\bibfnamefont{T.}~\bibnamefont{Feldmann}},
  \bibinfo{author}{\bibfnamefont{P.}~\bibnamefont{Kroll}}, \bibnamefont{and}
  \bibinfo{author}{\bibfnamefont{B.}~\bibnamefont{Stech}},
  \bibinfo{journal}{Phys. Rev.} \textbf{\bibinfo{volume}{D 58}},
  \bibinfo{pages}{114006} (\bibinfo{year}{1998}), \eprint{hep-ph/9802409}.

\bibitem[{\citenamefont{Feldmann}(2000)}]{Feldmann:1999uf}
\bibinfo{author}{\bibfnamefont{T.}~\bibnamefont{Feldmann}},
  \bibinfo{journal}{Int. J. Mod. Phys.} \textbf{\bibinfo{volume}{A 15}},
  \bibinfo{pages}{159} (\bibinfo{year}{2000}), \eprint{hep-ph/9907491}.

\bibitem[{\citenamefont{Acciarri et~al.}(1998)}]{Acciarri:1997yx}
\bibinfo{author}{\bibfnamefont{M.}~\bibnamefont{Acciarri}} \bibnamefont{et~al.}
  (\bibinfo{collaboration}{L3}), \bibinfo{journal}{Phys. Lett.}
  \textbf{\bibinfo{volume}{B 418}}, \bibinfo{pages}{399}
  (\bibinfo{year}{1998}).

\bibitem[{\citenamefont{Behrend et~al.}(1991)}]{Behrend:1990sr}
\bibinfo{author}{\bibfnamefont{H.~J.} \bibnamefont{Behrend}}
  \bibnamefont{et~al.} (\bibinfo{collaboration}{CELLO}), \bibinfo{journal}{Z.
  Phys.} \textbf{\bibinfo{volume}{C 49}}, \bibinfo{pages}{401}
  (\bibinfo{year}{1991}).

\bibitem[{\citenamefont{Gronberg et~al.}(1998)}]{Gronberg:1997fj}
\bibinfo{author}{\bibfnamefont{J.}~\bibnamefont{Gronberg}} \bibnamefont{et~al.}
  (\bibinfo{collaboration}{CLEO}), \bibinfo{journal}{Phys. Rev.}
  \textbf{\bibinfo{volume}{D 57}}, \bibinfo{pages}{33} (\bibinfo{year}{1998}),
  \eprint{hep-ex/9707031}.

\bibitem[{\citenamefont{del Amo~Sanchez et~al.}(2011)}]{BABAR:2011ad}
\bibinfo{author}{\bibfnamefont{P.}~\bibnamefont{del Amo~Sanchez}}
  \bibnamefont{et~al.} (\bibinfo{collaboration}{BaBar}),
  \bibinfo{journal}{Phys. Rev.} \textbf{\bibinfo{volume}{D 84}},
  \bibinfo{pages}{052001} (\bibinfo{year}{2011}), \eprint{1101.1142}.

\bibitem[{\citenamefont{Lees et~al.}(2018)}]{BaBar:2018zpn}
\bibinfo{author}{\bibfnamefont{J.~P.} \bibnamefont{Lees}} \bibnamefont{et~al.}
  (\bibinfo{collaboration}{BaBar}), \bibinfo{journal}{Phys. Rev.}
  \textbf{\bibinfo{volume}{D 98}}, \bibinfo{pages}{112002}
  (\bibinfo{year}{2018}), \eprint{1808.08038}.

\bibitem[{\citenamefont{Acciarri et~al.}(2001)}]{Acciarri:2000ev}
\bibinfo{author}{\bibfnamefont{M.}~\bibnamefont{Acciarri}} \bibnamefont{et~al.}
  (\bibinfo{collaboration}{L3}), \bibinfo{journal}{Phys. Lett.}
  \textbf{\bibinfo{volume}{B 501}}, \bibinfo{pages}{1} (\bibinfo{year}{2001}),
  \eprint{hep-ex/0011035}.

\bibitem[{\citenamefont{Ahohe et~al.}(2005)}]{Ahohe:2005ug}
\bibinfo{author}{\bibfnamefont{R.}~\bibnamefont{Ahohe}} \bibnamefont{et~al.}
  (\bibinfo{collaboration}{CLEO}), \bibinfo{journal}{Phys. Rev.}
  \textbf{\bibinfo{volume}{D 71}}, \bibinfo{pages}{072001}
  (\bibinfo{year}{2005}), \eprint{hep-ex/0501026}.

\bibitem[{\citenamefont{Ablikim et~al.}(2018)}]{Ablikim:2018ajr}
\bibinfo{author}{\bibfnamefont{M.}~\bibnamefont{Ablikim}} \bibnamefont{et~al.}
  (\bibinfo{collaboration}{BESIII}), \bibinfo{journal}{Phys. Rev.}
  \textbf{\bibinfo{volume}{D 97}}, \bibinfo{pages}{072014}
  (\bibinfo{year}{2018}), \eprint{1802.09854}.

\bibitem[{\citenamefont{Achard et~al.}(2007)}]{Achard:2007hm}
\bibinfo{author}{\bibfnamefont{P.}~\bibnamefont{Achard}} \bibnamefont{et~al.}
  (\bibinfo{collaboration}{L3}), \bibinfo{journal}{JHEP}
  \textbf{\bibinfo{volume}{03}}, \bibinfo{pages}{018} (\bibinfo{year}{2007}).

\bibitem[{\citenamefont{Zhang et~al.}(2012)}]{Zhang:2012tj}
\bibinfo{author}{\bibfnamefont{C.~C.} \bibnamefont{Zhang}} \bibnamefont{et~al.}
  (\bibinfo{collaboration}{Belle}), \bibinfo{journal}{Phys. Rev.}
  \textbf{\bibinfo{volume}{D 86}}, \bibinfo{pages}{052002}
  (\bibinfo{year}{2012}), \eprint{1206.5087}.

\bibitem[{\citenamefont{Klempt and Zaitsev}(2007)}]{Klempt:2007cp}
\bibinfo{author}{\bibfnamefont{E.}~\bibnamefont{Klempt}} \bibnamefont{and}
  \bibinfo{author}{\bibfnamefont{A.}~\bibnamefont{Zaitsev}},
  \bibinfo{journal}{Phys. Rept.} \textbf{\bibinfo{volume}{454}},
  \bibinfo{pages}{1} (\bibinfo{year}{2007}), \eprint{0708.4016}.

\bibitem[{\citenamefont{Sanchez-Puertas}(2019)}]{PabloSanchezPrivateCom}
\bibinfo{author}{\bibfnamefont{P.}~\bibnamefont{Sanchez-Puertas}},
  \bibinfo{journal}{private communications}  (\bibinfo{year}{2019}).

\bibitem[{\citenamefont{Knecht and Nyffeler}(2002)}]{Knecht:2001qf}
\bibinfo{author}{\bibfnamefont{M.}~\bibnamefont{Knecht}} \bibnamefont{and}
  \bibinfo{author}{\bibfnamefont{A.}~\bibnamefont{Nyffeler}},
  \bibinfo{journal}{Phys. Rev.} \textbf{\bibinfo{volume}{D 65}},
  \bibinfo{pages}{073034} (\bibinfo{year}{2002}), \eprint{hep-ph/0111058}.

\bibitem[{\citenamefont{Prades et~al.}(2009)\citenamefont{Prades, de~Rafael,
  and Vainshtein}}]{Prades:2009tw}
\bibinfo{author}{\bibfnamefont{J.}~\bibnamefont{Prades}},
  \bibinfo{author}{\bibfnamefont{E.}~\bibnamefont{de~Rafael}},
  \bibnamefont{and}
  \bibinfo{author}{\bibfnamefont{A.}~\bibnamefont{Vainshtein}},
  \bibinfo{journal}{Adv. Ser. Direct. High Energy Phys.}
  \textbf{\bibinfo{volume}{20}}, \bibinfo{pages}{303} (\bibinfo{year}{2009}),
  \eprint{0901.0306}.

\bibitem[{\citenamefont{Davier et~al.}(2008)\citenamefont{Davier,
  Descotes-Genon, Hocker, Malaescu, and Zhang}}]{Davier:2008sk}
\bibinfo{author}{\bibfnamefont{M.}~\bibnamefont{Davier}},
  \bibinfo{author}{\bibfnamefont{S.}~\bibnamefont{Descotes-Genon}},
  \bibinfo{author}{\bibfnamefont{A.}~\bibnamefont{Hocker}},
  \bibinfo{author}{\bibfnamefont{B.}~\bibnamefont{Malaescu}}, \bibnamefont{and}
  \bibinfo{author}{\bibfnamefont{Z.}~\bibnamefont{Zhang}},
  \bibinfo{journal}{Eur. Phys. J.} \textbf{\bibinfo{volume}{C 56}},
  \bibinfo{pages}{305} (\bibinfo{year}{2008}), \eprint{0803.0979}.

\bibitem[{\citenamefont{Beneke and Jamin}(2008)}]{Beneke:2008ad}
\bibinfo{author}{\bibfnamefont{M.}~\bibnamefont{Beneke}} \bibnamefont{and}
  \bibinfo{author}{\bibfnamefont{M.}~\bibnamefont{Jamin}},
  \bibinfo{journal}{JHEP} \textbf{\bibinfo{volume}{09}}, \bibinfo{pages}{044}
  (\bibinfo{year}{2008}), \eprint{0806.3156}.

\bibitem[{\citenamefont{Maltman and Yavin}(2008)}]{Maltman:2008nf}
\bibinfo{author}{\bibfnamefont{K.}~\bibnamefont{Maltman}} \bibnamefont{and}
  \bibinfo{author}{\bibfnamefont{T.}~\bibnamefont{Yavin}},
  \bibinfo{journal}{Phys. Rev.} \textbf{\bibinfo{volume}{D 78}},
  \bibinfo{pages}{094020} (\bibinfo{year}{2008}), \eprint{0807.0650}.

\bibitem[{\citenamefont{Narison}(2009)}]{Narison:2009vy}
\bibinfo{author}{\bibfnamefont{S.}~\bibnamefont{Narison}},
  \bibinfo{journal}{Phys. Lett.} \textbf{\bibinfo{volume}{B 673}},
  \bibinfo{pages}{30} (\bibinfo{year}{2009}), \eprint{0901.3823}.

\bibitem[{\citenamefont{Caprini and Fischer}(2009)}]{Caprini:2009vf}
\bibinfo{author}{\bibfnamefont{I.}~\bibnamefont{Caprini}} \bibnamefont{and}
  \bibinfo{author}{\bibfnamefont{J.}~\bibnamefont{Fischer}},
  \bibinfo{journal}{Eur. Phys. J.} \textbf{\bibinfo{volume}{C 64}},
  \bibinfo{pages}{35} (\bibinfo{year}{2009}), \eprint{0906.5211}.

\bibitem[{\citenamefont{Pich}(2014)}]{Pich:2013lsa}
\bibinfo{author}{\bibfnamefont{A.}~\bibnamefont{Pich}}, \bibinfo{journal}{Prog.
  Part. Nucl. Phys.} \textbf{\bibinfo{volume}{75}}, \bibinfo{pages}{41}
  (\bibinfo{year}{2014}), \eprint{1310.7922}.

\bibitem[{\citenamefont{Leutgeb and Rebhan}(2019)}]{Leutgeb:2019gbz}
\bibinfo{author}{\bibfnamefont{J.}~\bibnamefont{Leutgeb}} \bibnamefont{and}
  \bibinfo{author}{\bibfnamefont{A.}~\bibnamefont{Rebhan}}
  (\bibinfo{year}{2019}), \eprint{1912.01596}.

\bibitem[{\citenamefont{Cappiello et~al.}(2019)\citenamefont{Cappiello,
  Cat{\`a}, D'Ambrosio, Greynat, and Iyer}}]{Cappiello:2019hwh}
\bibinfo{author}{\bibfnamefont{L.}~\bibnamefont{Cappiello}},
  \bibinfo{author}{\bibfnamefont{O.}~\bibnamefont{Cat{\`a}}},
  \bibinfo{author}{\bibfnamefont{G.}~\bibnamefont{D'Ambrosio}},
  \bibinfo{author}{\bibfnamefont{D.}~\bibnamefont{Greynat}}, \bibnamefont{and}
  \bibinfo{author}{\bibfnamefont{A.}~\bibnamefont{Iyer}}
  (\bibinfo{year}{2019}), \eprint{1912.02779}.

\end{thebibliography}

\end{document}